\documentclass[lettersize,journal]{IEEEtran}
\usepackage{amsmath,amsfonts}
\usepackage{algpseudocode}
\usepackage{algorithm}
\usepackage{array}
\usepackage[caption=false,font=footnotesize,labelfont=normalfont,textfont=normalfont]{subfig}
\usepackage{textcomp}
\usepackage{stfloats}
\usepackage{url}
\usepackage{amsmath}
\usepackage{verbatim}
\usepackage{graphicx}
\usepackage{hyperref}
\usepackage{cite}
\usepackage{todonotes}
\begin{document} 

\title{Vision and Causal Learning Based Channel Estimation for THz Communications}

\author{Kitae~Kim,~\IEEEmembership{Member,~IEEE},~Yan~Kyaw~Tun,~\IEEEmembership{Member,~IEEE},~Md.~Shirajum~Munir,~\IEEEmembership{Member,~IEEE},~Christo Kurisummoottil Thomas,~\IEEEmembership{ Member,~IEEE,}~Walid~Saad,~\IEEEmembership{Fellow,~IEEE,}~and~Choong~Seon~Hong,~\IEEEmembership{Fellow,~IEEE}
\thanks{Kitae~Kim and Choong Seon Hong are with the Department of Computer Science and Engineering, Kyung Hee University,  Yongin-si, Gyeonggi-do 17104, Rep. of Korea, e-mails:{\{glideslope, cshong\}@khu.ac.kr.}}
\thanks{Yan Kyaw Tun is with the Department of Electronic Systems, Aalborg University, 2450 København SV, Denmark, e-mail:ykt@es.aau.dk.}
\thanks{Md. Shirajum Munir is with the School of Computing, Analytics, and Modeling, University of West Georgia, Carrollton, GA 30118, USA, e-mail: mmunir@westga.edu.}
\thanks{Christo Kurisummoottil Tomas and Walid Saad are with the Department of Electrical and Computer Engineering, Virginia Tech, Arlington, VA 22203, USA, e-mails: \{christokt, walids\}@vt.edu.} }

\markboth{Journal of \LaTeX\ Class Files,~Vol.~14, No.~8, August~2021}%
{Shell \MakeLowercase{\textit{et al.}}: A Sample Article Using IEEEtran.cls for IEEE Journals}

\maketitle
\begin{abstract}
The use of terahertz (THz) communications with massive multiple input multiple output (MIMO) systems in 6G can potentially provide high data rates and low latency communications. However, accurate channel estimation in THz frequencies presents significant challenges due to factors such as high propagation losses, sensitivity to environmental obstructions, and strong atmospheric absorption. These challenges are particularly pronounced in urban environments, where traditional channel estimation methods often fail to deliver reliable results, particularly in complex non-line-of-sight (NLoS) scenarios. This paper introduces a novel vision-based channel estimation technique that integrates causal reasoning into urban THz communication systems. The proposed method combines computer vision algorithms with variational causal dynamics (VCD) to analyze real-time images of the urban environment, allowing for a deeper understanding of the physical factors that influence THz signal propagation. By capturing the complex, dynamic interactions between physical objects (such as buildings, trees, and vehicles) and the transmitted signals, the model can predict the channel with up to twice the accuracy of conventional methods. This model improves estimation accuracy and demonstrates superior generalization performance. Hence, it can provide reliable predictions even in previously unseen urban environments. The effectiveness of the proposed method is particularly evident in NLoS conditions, where it significantly outperforms traditional methods such as by accounting for indirect signal paths, such as reflections and diffractions. Simulation results confirm that the proposed vision-based approach surpasses conventional artificial intelligence (AI)-based estimation techniques in accuracy and robustness, showing a substantial improvement across various dynamic urban scenarios. This framework provides a promising solution for enabling resilient THz communication, offering scalability and practicality for future 6G deployments in diverse urban landscapes. 
\end{abstract}


\section{Introduction}
\IEEEPARstart{T}{he} advancement of 6G wireless networks is poised to significantly enhance next-generation wireless communication by enabling ultra-high bandwidth and low-latency services, crucial for data-intensive applications such as virtual reality, autonomous driving, and the Internet of Everything (IoE) \cite{8869705,9509294,9310315}. Thus, the increased demand for higher densities of connected devices and support for rapid data transmission over vast volumes have become increasingly essential in 6G networks. To address these requirements, there is a need for new approaches to ensure reliable and efficient communication in complex and rapidly changing environments.

One of the most promising techniques is the integration of terahertz (THz) frequencies into current wireless networks, as the THz band (0.1–10 THz) offers substantially greater bandwidth, facilitating the ultra-high-speed data rates envisioned for 6G\cite{9782674,9681870}. However, utilizing THz frequencies introduces a number of challenges due to the unique propagation characteristics of THz signals, which experience high propagation losses and are sensitive to physical obstructions and atmospheric absorption \cite{10579941}. In particular, THz waves are susceptible to absorption by atmospheric molecules, such as water vapor, resulting in significant signal attenuation over longer distances. Additionally, the shorter wavelengths associated with THz frequencies struggle to penetrate solid materials, making them particularly vulnerable to blockages from urban infrastructure\cite{9845935}. Consequently, THz communication inevitably relies heavily on line-of-sight (LoS) propagation, which makes signal transmission particularly difficult in non-LoS environments. This challenge is especially significant in dense urban areas. To mitigate these challenges, advanced multiple input multiple output (MIMO) systems play a crucial role in THz communications\cite{10078317,10379539}. MIMO uses multiple transmit and receive antennas to enhance spatial diversity and enable beamforming, which helps improve signal strength and reduce blockage issues. In particular, beamforming at THz frequencies enables the formation of narrow beams, increasing signal directivity and allowing adaptive path selection, thereby improving reliability even in NLoS conditions. However, leveraging MIMO in THz systems effectively requires accurate channel state information (CSI).

The main contribution of this paper is a novel channel estimation framework that leverages computer vision and causal inference to extract environmental features and model their impact on THz signal propagation. By utilizing real-time urban environment images and variational causal dynamics (VCD) \cite{10096153,lei2022variational}, the proposed method estimates the direct impact of physical obstructions on signal paths while identifying the underlying causal relationships that influence THz signal propagation. Specifically, the proposed approach integrates computer vision and causal inference to create a model that dynamically adapts to the changing features of urban environments. Using images from diverse vantage points, the model identifies key environmental characteristics such as building materials, vehicular mobility, etc. that impact THz signal propagation.

These identified inputs are then processed within a VCD framework, which leverages domain knowledge to define an initial causal structure—an essential step since purely data-driven methods often fail to infer physically consistent relationships in complex wireless settings. By integrating structured causal modeling (SCM) with variational inference, the framework refines these relationships rather than discovering entirely new physical laws, thus preserving well-established propagation principles while adapting to scenario changes. As a result, the model more effectively captures the interplay between physical obstructions and signal behavior leading to accurate and reliable channel estimations even in challenging and rapidly evolving urban environments.

To the best of our knowledge, this is the first attempt at performing \emph{channel estimation through causal modeling of the relationship between channel variables and environmental factors.} Rather than constraining estimation to a pre-defined analytic or probabilistic channel model, we learn a model-agnostic mapping that links environment to channel variables and remains valid across mathematical, probabilistic, and measurement-based formulations. Concretely, given any channel model instantiated for a propagation environment, our objective is to first identify how its constituent variables change with variations in the environment or other operating conditions, and then to use these identified relations to accurately estimate and predict the channel. This perspective is particularly important for THz communications, where signal propagation is highly sensitive to material properties, blockage, sparsity, and environmental dynamics. By uncovering causal relations rather than relying on correlations, the proposed framework enhances interpretability, robustness, and generalization for channel estimation in the THz domain.
 In summary, our key contributions include:
\begin{itemize}
    \item We study how to leverage causal variables for channel estimation that capture the interactions between channel model parameters (e.g., path gain, AoA, AoD) and environmental factors (e.g., object size, material properties, spatial positioning) in urban THz wireless environments.

    \item We introduce a novel vision-based approach that integrates RGB-D camera inputs with a VCD to infer causal relationships between urban environmental features and THz channel characteristics. This approach explicitly models the impact of obstacles and multipath reflections, significantly improving channel parameter estimation.

    \item We then develop a dynamic environment adaptation framework that updates causal structures in response to environmental changes, enabling robust generalization to unseen urban layouts. Unlike conventional deep learning models that require retraining, 
    Our approach leverages variational inference to adapt efficiently to new scenarios.

    \item Through extensive simulations using MATLAB Ray Tracing and CARLA-generated environments, we demonstrate that the proposed approach outperforms conventional AI-based channel estimation methods without a causality mechanism in both accuracy and generalization ability. Our method achieves up to a 6.6 times improvement in channel estimation performance compared to traditional deep learning approaches and maintains robust performance even in unseen dynamic urban scenarios.
\end{itemize}

The rest of this paper is organized as follows. Section II reviews related work. Section III describes the system model. Solution approaches are presented in Section IV. Simulation results are provided in Section V. Finally, conclusions are drawn in Section VI.

\section{Related Work}
While conventional channel estimation techniques perform well at lower frequencies, they cannot adapt to THz communication due to their unique characteristics, such as high path loss, signal attenuation, and extreme sensitivity to environmental changes. Therefore, it is essential to develop robust THz channel modeling and estimation methods to handle frequent physical obstructions and varying atmospheric conditions. To address this issue, prior works \cite{8640815,8949757,10461569,10143629,9452036} have explored deep neural network-based channel estimation techniques as a promising approach. A deep learning-based approach can help extract complex features from data and provide more accurate channel estimation under dynamic environmental changes.

Deep learning-based methods offer clear advantages, but existing statistical deep learning approaches are strongly dependent on specific datasets. A neural network trained on data collected in a particular environment often fails to generalize to new environments. As a result, whenever a channel estimator is used in a different setting, the model must be retrained, leading to additional costs. Since channel estimation must be performed across various locations and conditions, this issue becomes even more critical. To address this, continual learning (CL) was used \cite{10444954} and \cite{10447575} to enable models to adapt to new environments without requiring full retraining. However, CL-based methods may not be well-suited for THz channel estimation, where wireless conditions change rapidly over time and space. CL cannot effectively handle sudden variations in the channel, and trained models may either rely too much on past data or fail to incorporate new information effectively. Therefore, a more adaptive channel estimation method is required to ensure reliable performance across different environments. Overall, existing studies rely on correlation-driven mappings and narrow training domains, which limit robustness under material-sensitive propagation and rapid environmental shifts in THz links. We therefore present a vision-based, causality-aware framework that addresses these limitations.

\section{System Model}
\begin{figure}[ht]
	\centering
	\captionsetup{justification=centering}
	\includegraphics[width=8.8cm, height=5.5cm]{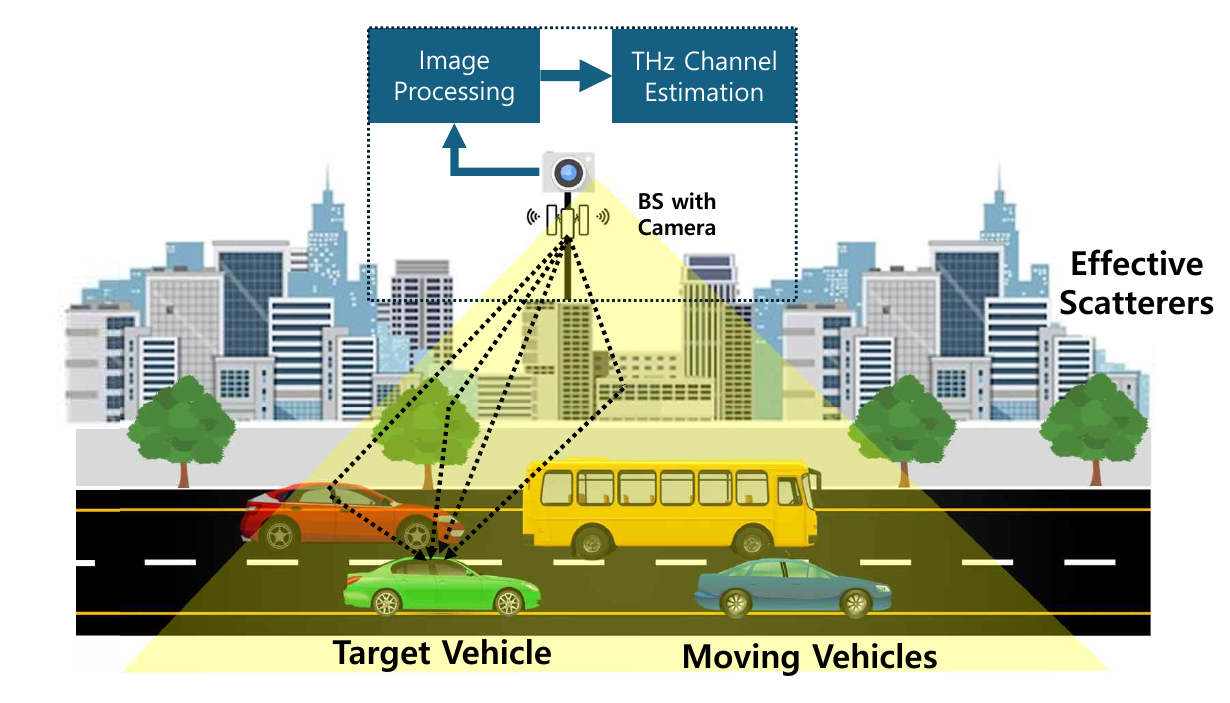}
	\caption{System model of the proposed framework.}
	\label{systemmodel}
\end{figure}

We consider a downlink vehicle-to-infrastructure (V2I) communication scenario, illustrated in Fig.~\ref{systemmodel}, where a base station (BS) equipped with \( N_t \) uniform linear array (ULA) antennas provides communication links to the set \(\mathcal{I}\) of \( I \) users, each equipped with \( N_r \) ULA antennas. Furthermore, \( N_c \) RGB-depth cameras are deployed at the BS, each positioned at different azimuth angles to capture multiview images. Meanwhile, a
set \(\mathcal{J}\) of \( J \) surrounding objects is also present in the considered environment. These objects can cause reflections, diffractions, and blockages of the signals, influencing the communication channel. All the cameras are assumed to be positioned at the exact location corresponding to the center of the antenna array. According to \cite{10107634}, for practical scenarios in which the cameras are positioned at a distance from the antenna, advanced view synthesis techniques can be employed to create images that replicate the perspective from the antenna's viewpoint using images from other camera angles.

\subsection{Communication Model}
We adopt the Saleh-Valenzuela model for the THz channel as described in \cite{10185655} and \cite{9514889}. Thus, the channel matrix \(\boldsymbol{H}_k\) at time index \(k\) is given by:
\begin{equation}\label{eqn:channelmodel}
\begin{split}
    \boldsymbol{H}_k = \gamma^{(0)}_k \eta(d^{(0)}_k) \boldsymbol{a}_r(\varphi^{(0)}_{r,k}) \boldsymbol{a}^H_t (\varphi^{(0)}_{t,k}) \\ 
    + \sum_{l=1}^{L} \gamma^{(l)}_k \eta(d^{(l)}_k) \boldsymbol{a}_r(\varphi^{(l)}_{r,k}) \boldsymbol{a}^H_t (\varphi^{(l)}_{t,k}),
\end{split}
\end{equation}
where \(L\) is the number of reflection paths. Function $\eta(\cdot)$ represents the path gain, while \(d^{(0)}_k\) and \(d^{(l)}_k\) represent the lengths of the LoS path and the \(l\)-th reflection path, respectively. The angles of departure (AoD) and arrival (AoA) for path $l$ are given by \(\varphi^{(l)}_t\) and \(\varphi^{(l)}_r\), respectively. Vectors \(\boldsymbol{a}_t(\cdot)\) and \(\boldsymbol{a}_r(\cdot)\) represent the transmit and receive array response vectors, respectively. Furthermore, \(\boldsymbol{a}_t^H(\cdot)\) represents the hermitian transpose of the transmit array response vector, which ensures proper alignment of the transmitted and received signal components in the channel model. To account for blockage, we define a binary variable, \(\gamma^{(l)}_k \in \{0,1\} \), which is equal to \(1\) if link $l$ exists at time step \(k\) and \(0\) otherwise. For simplicity, we assume the antenna spacing to be half a wavelength. Therefore, the array response vectors are given by:
\begin{equation}
\boldsymbol{a}(\varphi) = \frac{1}{\sqrt{N}} \begin{bmatrix}
1, & e^{j\pi \sin(\varphi)}, & \ldots, & e^{j\pi(N-1) \sin(\varphi)}
\end{bmatrix}^T,
\end{equation}
where \(N\) is the number of antenna elements in the array. A larger \(N\) enhances the resolution and spatial visibility of the array, enabling more precise beamforming and direction finding. For simplicity, we assume the antenna spacing to be half a wavelength.
At the THz frequency bands, molecular absorption greatly impacts path loss and free space propagation loss. As a result, the overall channel gain will be given by \cite{9678373}:
\begin{equation}
\eta(d_k) = \frac{c}{4\pi f d_k} e^{-\frac{1}{2} K(f) d_k},
\end{equation}
where \(f\) is the carrier frequency, \(c\) is the speed of light, and \(K(f)\) is the overall absorption coefficient of the medium. Therefore, the received signal at user \(k\) will be:
\begin{equation}
y_k = w_k^H \boldsymbol{H}_k \boldsymbol{f}_k s_k + n_k,
\end{equation}
where $\boldsymbol{f}_k$, and $w_k$ represent the precoder and combiner, respectively, while $n_k \sim \mathcal{CN}(0, \sigma^2_n)$ represents the additive white Gaussian noise (AWGN). The noise power $\sigma^2_n$ is given by:
\begin{equation}
\sigma^2_n = N_0 + \left( \frac{P_{\text{max}}}{4 \pi f d_k} \right)^2 (1 - e^{-K(f) d_k}),
\end{equation}
Here, $N_0 = \frac{W \lambda^2}{4 \pi k_B T_0}$ represents the thermal noise power, where $k_B$ is the Boltzmann constant, $W$ is the bandwidth and $T_0$ is the temperature in Kelvin. $\lambda$ represents the wavelength of the signal, defined as $\lambda = \frac{c}{f}$, $f$ being the frequency and $c$ being the speed of light. Moreover, the noise also accounts for molecular absorption, which arises due to molecular re-radiation \cite{9678373}.
Given the prominence of the LoS in the THz channel, the precoder and combiner should ideally be designed to direct the beam along the LoS path. If the LoS path is blocked, they should align with the strongest reflection path. Because the THz frequency bands use narrow beams, the power received from reflections and scattering at angles other than the main lobe direction is negligible \cite{miretti2024little}. Therefore, interference from other devices is also ignored. Consequently, since only the strongest path is used for transmission, we approximate the channel as follows:
\begin{equation}\label{eqn:approximate}
 \boldsymbol{H}_k \approx \gamma_k \eta(d_k) \boldsymbol{a}_r(\phi_{r,k}) \boldsymbol{a}_t^H(\phi_{t,k}),
\end{equation}
where $d_k$, $\phi_{t,k}$, and $\phi_{r,k}$ represent the distance, the AoD, and the AoA of the LoS link, or of the strongest available NLoS path, respectively. Note that our goal is to uncover causal links between environmental and channel variables rather than to fit a specific channel model. Thus, the approach is model-agnostic and can be applied to various models in both real and simulated settings. Representative 5G/6G and THz geometry-based channel models include 3GPP TR~38.901~\cite{3gpp38901}, IEEE~802.15.3d~\cite{IEEE802153d2017}. In addition, measurement-based channel simulators such as NYUSIM~\cite{NYUSIM2017} provide angles, delays, path powers, and blockage or mobility variables. These variables can be used within the same causal-learning framework, and this shows that the approach is not tied to a single model formulation.

\subsection{Environmental Features}
Obstacles significantly affect the propagation of THz signals in urban environments, necessitating the precise extraction of these elements for reliable channel estimation. The objective is to use RGB-D images to capture critical environmental features and establish causal relationships with wireless channel variables. RGB-D images provide both RGB color and depth information, offering a visual representation of positional, dimensional, and material properties of environmental factors that influence signal propagation.
We extract environmental features, denoted by $\boldsymbol{F}$, from both the target and the surrounding objects as follows:
\begin{equation}
\boldsymbol{F} = \{\boldsymbol{F}_t, \boldsymbol{F}_s\},
\end{equation}
where $\boldsymbol{F}_t$ represents the features of the target object (i.e., the UE), and $\boldsymbol{F}_s$ encompasses the features of multiple surrounding objects.
Since there is only one target object, we define its feature vector as:
\begin{equation}
\label{eq:Ft}
\boldsymbol{F}_t 
= 
\bigl[x_t,\; y_t,\; z_t,\; r_t,\; \text{AoA}_t,\; \text{AoD}_t\bigr].
\end{equation}
Here, $(x_t, y_t, z_t)$ denotes the 3D coordinates of the UE, while $r_t$ represents the distance between the UE and the transmitter. The terms $\text{AoA}_t$ and $\text{AoD}_t$ are the angles at which the signal arrives and departs from the UE, respectively.
In contrast, $\boldsymbol{F}_s$ represents the features of $J$ surrounding objects that influence channel estimation through reflection, scattering, and obstruction:
\begin{equation}
\boldsymbol{F}_s = \{\boldsymbol{F}_{s,j} \mid j = 1, \dots, J\}.
\end{equation}
Each surrounding object $j$ contributes a feature vector
\begin{equation}
\label{eq:Fsj}
\boldsymbol{F}_{s,j} = 
\bigl[x_j,\; y_j,\; z_j,\; w_j,\; h_j,\; d_j,\; m_j,\; r_j,\; \text{AoA}_j,\; \text{AoD}_j\bigr].
\end{equation}
The terms $(x_j, y_j, z_j)$ specify the 3D coordinates of the object $j$, while $(w_j, h_j, d_j)$ describe its physical dimensions in terms of width, height, and depth, respectively. The variable $m_j$ characterizes the object's material properties, and $r_j$ denotes its spatial distance from the transmitter or receiver. Lastly, $(\text{AoA}_j, \text{AoD}_j)$ define the angles of arrival and departure of the signal with respect to the object, which are crucial for modeling reflection and scattering effects. This detailed feature set $\boldsymbol{F}$ comprehensively represents the urban environment by leveraging RGB-D camera images to capture both the target's and the surrounding objects' spatial and material properties.

\section{Causal Variables for THz Channel}  
Having defined the environmental factors influencing channel variations, we next identify the key variables essential for modeling the causal relationships between these environmental changes and the channel parameters. The variables are defined to systematically represent the interactions between the environment and the THz communication channel. The channel parameters $\boldsymbol{X}$ are defined as follows:

\begin{equation}
\boldsymbol{X} = [\boldsymbol{X_1}, \boldsymbol{X_2}, \boldsymbol{X_3}].
\end{equation}

\begin{itemize}
    \item $\boldsymbol{X}_1$ (Path gain, \( \gamma^{(l)}_k \)): Influenced by the distances (\( r_t, r_j \)), material properties (\( m_j \)), and reflective or diffractive characteristics of the environment.
    \item $\boldsymbol{X}_2$ (Angles, \( \phi^{(l)}_{r,k}, \phi^{(l)}_{t,k} \)): Determined by the angular relationships of the target (\( \phi_{r,t}, \phi_{t,t} \)) and the positions, dimensions, and velocities of surrounding objects (\( (x_j, y_j, z_j), (w_j, h_j, d_j), v_j \)).
    \item $\boldsymbol{X}_3$ (Path distance, \( d^{(l)}_k \)): Affected by the 3D positions of the target (\( x_t, y_t, z_t \)) and surrounding objects (\( x_j, y_j, z_j \)), as well as the distances to these objects (\( r_t, r_j \)).
\end{itemize}

The environmental features $\boldsymbol{F}$ are divided into target features $\boldsymbol{F}_t$ and surrounding features $\boldsymbol{F}_s$:

\begin{itemize}
    \item Target features ($ \boldsymbol{F_t} $): These describe the characteristics of the user equipment (UE) and include:
        $ \boldsymbol{E}_1 $ (3D position of the UE, $ x_t, y_t, z_t $), 
        $ E_2 $ (distance from the transmitter, $ r_t $), and 
        $ \boldsymbol{E}_3 $ (angular relationships for AoA and AoD, $ \phi_{r,t}, \phi_{t,t} $).
    
    \item Surrounding features ($ \boldsymbol{F_s} $): These represent the properties of objects in the environment that influence the channel, including:
        $ \boldsymbol{E}_4 $ (3D positions of surrounding objects, $ x_j, y_j, z_j $), 
        $ \boldsymbol{E}_5 $ (dimensions of objects, $ w_j, h_j, d_j $), 
        $ E_6 $ (material properties of objects, $ m_j $), 
        $ \boldsymbol{E}_7 $ (velocities of dynamic objects, $ v_j $), 
        $ E_8 $ (distances from the transmitter or receiver, $ r_j $), and 
        $ \boldsymbol{E}_9 $ (angular relationships for AoA and AoD, $ \phi_{r,j}, \phi_{t,j} $).
\end{itemize}
These environmental factors provide a structured representation of the relationship between the environment and the channel parameters, which enables accurate causal modeling.

Recent studies on causal representation learning highlight that discovering high-level causal variables from raw data remains a fundamental challenge. As noted in \cite{9363924}, machine learning models excel at capturing statistical dependencies but struggle to infer physically meaningful causal structures without domain knowledge. In particular, deep learning excels at learning representations of data that preserve relevant statistical properties. However, it does so without taking into account the causal properties of the variables. This underscores the importance of incorporating physically grounded variables into the model rather than relying solely on data-driven discovery.

To address this limitation, we explicitly define the environmental variables $ \boldsymbol{F} $ and channel parameters $ \boldsymbol{X} $ based on well-established physical principles governing THz wave propagation. Accordingly, we inject lightweight THz prior knowledge into the causal graph. In particular, distance governs path length and attenuation~\cite{7504435,9558848}, object size and material modulate the resulting path gain~\cite{9013236,8761205}, angles of departure and arrival determine whether the reflected energy aligns with the TX/RX beams~\cite{7289335}, and human/vehicle blockage controls whether a path exists at all~\cite{7881087,10896959}. We encode these priors as initial edges and allow only sparse refinements during learning, which improves training stability and generalization. This approach ensures that the model learns interpretable and generalizable relationships rather than statistical correlations. By integrating domain knowledge, we improve learning efficiency, enhance robustness to environmental variations, and align our causal framework with real-world physical constraints. The relationships between $ \boldsymbol{F} $, $ \boldsymbol{E} $, and the channel parameters ($ \boldsymbol{X}_1, \boldsymbol{X}_2, \boldsymbol{X}_3 $) are governed by well-defined physical principles. Static structures $ \boldsymbol{E}_4 $, $ \boldsymbol{E}_5 $, $ \boldsymbol{E}_6 $ influence path gain $ \boldsymbol{X}_1 $ and signal angles $ \boldsymbol{X}_2 $ through reflection and diffraction. Dynamic elements $ \boldsymbol{E}_7 $, $ \boldsymbol{E}_9 $ cause temporal variations in the channel by altering signal paths due to object velocities and changes in angular relationships.  
Explicitly defining the causal relationships between environmental factors and channel characteristics is essential to enhance the reliability and interpretability of the model. However, to fully exploit the advantages of these defined variables, an effective modeling approach capable of efficiently extracting environmental features and learning causal relationships is required. Therefore in the following section we propose a detailed methodology that effectively extracts the defined causal variables from visual information and leverages them to model causal relationships with the channel parameters to enable accurate estimation of the THz channel.


\begin{figure*}
    \centering
    \includegraphics[width=0.85\textwidth]{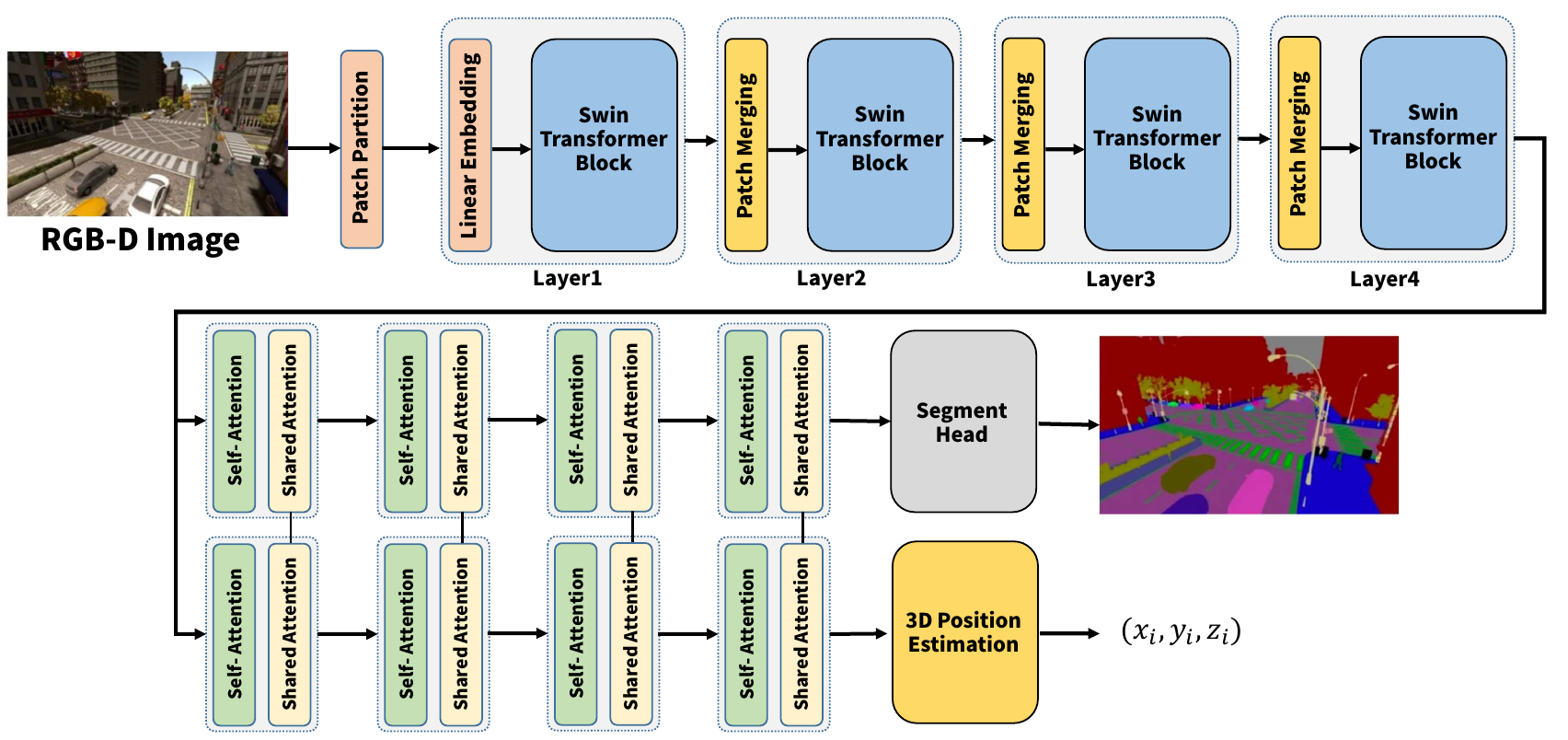}
    \caption{Multi-task learning framework for environmental feature extraction.}
    \label{fig:mt}
\end{figure*}

\section{Vision Information and Causal Learning based Channel Estimation }
This section presents a vision-assisted solution to estimate the channel matrix $\boldsymbol{H}_k$ at time $k$ in THz communications. Our proposed approach leverages explicitly defined causal relationships between environmental features and channel parameters by integrating the multitask transformer (MulT) model \cite{Bhattacharjee_2022_CVPR} and a causal neural network. Specifically, the MulT model first processes RGB-D images to extract essential environmental variables, such as object positions, dimensions, and material properties, as defined in Section II. The causal neural network then learns the causal relationships between these extracted environmental variables and the predefined channel parameters also described in Section II. In the following, we detail how these components interact to achieve an accurate and physically consistent estimation of the THz channel.

\subsection{MulT for Environmental Feature Extraction}
We now explain how the MulT model is employed to extract features of target objects and surrounding objects, while focusing on 3D coordinate estimation and environmental segmentation. The MulT model utilizes a Swin Transformer-based encoder-decoder architecture to extract spatial and semantic features from RGB-D images. The primary tasks include semantic segmentation, which classifies each pixel such as buildings, trees, or vehicles, and 3D coordinate prediction, which estimates the physical coordinates \((\hat{x}, \hat{y}, \hat{z})\) for each pixel. These tasks are trained using RGB-D input data and ground truth from Carla \cite{Dosovitskiy17} to enable precise predictions.
RGB-D data comprises RGB channels for texture and boundary information and depth channels for spatial structure. The encoder processes this data using the shifted window self-attention mechanism to learn spatial and semantic relationships. The RGB channels help identify objects' textures and boundaries, while the depth channels provide critical spatial information, such as distances between objects and the camera. This combination allows the encoder to capture both local details and global structure in the environment.
The encoder learns features to support the two primary tasks. The encoder learns semantic relationships for semantic segmentation by analyzing the RGB data for texture, color, and boundary features. For instance, it distinguishes between the textures of building surfaces and tree leaves, grouping pixels belonging to the same object. For 3D coordinate prediction, the encoder learns spatial relationships by leveraging depth data and the relative positions of pixels. This learning helps model spatial continuity and differentiate distances between objects.

The self-attention mechanism in the encoder enables learning interactions between pixels by calculating attention scores. These scores are defined as \cite{vaswani2017attention}:
\begin{equation}
A = \text{softmax}\left(\frac{QK^T}{\sqrt{d_k}} + B\right)V,
\end{equation}
where \(Q, K\), and \(V\) represent query, key, and value matrices derived from pixel features, \(d_k\) is the dimensionality of the key, and \(B\) is a positional bias that encodes spatial relationships. For semantic segmentation, the mechanism highlights texture and boundary similarities, grouping pixels into consistent categories. For 3D coordinate prediction, it captures depth continuity and relative spatial positions.

Shifted window self-attention divides the input into fixed windows and computes the attention within each window. Shifting these windows across layers ensures interaction across window boundaries. This, in turn, allows the model to capture details of small objects and the relationships between larger structures. The encoder generates hierarchical feature maps at different resolutions:
\begin{equation}
\text{Resolution: } \frac{H_{img}}{2^l} \times \frac{W_{img}}{2^l}, \quad \text{Channels: } 2^l \cdot C,
\end{equation}
where $H_{img}$, and $W_{img}$ are input dimensions, $l$ is the layer index, and $C$ is the base channel size. Lower resolution maps capture global layouts and large objects, while higher resolution maps retain fine-grained details.

The MulT model uses shared attention in the decoder to integrate the hierarchical features of the encoder and adapt them to the requirements of the two tasks. Shared Attention aligns encoder features $(x_{sa}$ with task-specific inputs $x_t$, and attention is calculated as:
\begin{equation}
A_{sa}^r = \text{softmax}\left(\frac{Q^r (K^r)^T}{\sqrt{d_k}} + B^r\right),
\end{equation}
where \(Q^r\) and \(K^r\) are task-specific query and key matrices, and \(B^r\) represents task-specific positional biases. Shared Attention produces refined features for each task, represented as:
\begin{equation}
\tilde{x}_t = A_{sa}^r V^t
\end{equation}
where \(V^t\) is the task-specific value matrix. This output is passed to independent decoder heads for the two tasks.

For 3D coordinate prediction, the decoder estimates the spatial location of each pixel:
\begin{equation}
(\hat{x}, \hat{y}, \hat{z}) = W_{\text{pos}} x_t,
\end{equation}
where \(W_{\text{pos}}\) is the learned weight for spatial regression. The decoder ensures spatial continuity within objects and distinguishes objects based on depth differences.

For semantic segmentation, the decoder classifies each pixel into predefined categories:
\begin{equation}
\hat{y}_{\text{seg}} = \text{softmax}(W_{\text{seg}} x_t),
\end{equation}
where \(W_{\text{seg}}\) represents task-specific weights. This process maintains consistent classification across object boundaries by leveraging texture and depth information.

The MulT model optimizes these predictions using a joint loss function. The 3D coordinate regression loss minimizes the Euclidean distance between predicted and actual coordinates:
\begin{equation}
\mathcal{L}_{\text{pos}} = \frac{1}{|\mathcal{J}|} \sum_{j \in \mathcal{J}} \left((x_j - \hat{x}_j)^2 + (y_j - \hat{y}_j)^2 + (z_j - \hat{z}_j)^2\right),
\end{equation}
The semantic segmentation loss evaluates pixel classification accuracy using cross-entropy:
\begin{equation}
\mathcal{L}_{\text{seg}} = -\frac{1}{N} \sum_{p=1}^N \sum_{c=1}^C y_{p,c} \log(\hat{y}_{p,c}),
\end{equation}
where \( N \) is the total number of points (e.g., image pixels or spatial locations) in the image or dataset being evaluated, and \( C \) is the total number of classes for segmentation. \( y_{p,c} \) represents the ground truth label for point \( p \) and class \( c \), where \( y_{p,c} = 1 \) if point \( p \) belongs to class \( c \), and \( y_{p,c} = 0 \) otherwise. \( \hat{y}_{p,c} \) represents the predicted probability that point \( p \) belongs to class \( c \), and \( \log(\hat{y}_{p,c}) \) is the logarithm of this predicted probability. The loss \( \mathcal{L}_{\text{seg}} \) calculates the average cross-entropy over all points and all classes.
The total loss balances the two tasks:
\begin{equation}
\mathcal{L}_{\text{total}} = \alpha \mathcal{L}_{\text{seg}} + \beta \mathcal{L}_{\text{pos}},
\end{equation}
where \(\alpha\), and \(\beta\) are weights controlling the importance of each task.
The encoder learns shared features, such as object boundaries and depth continuity, which are critical for both tasks. Shared Attention ensures both decoders effectively utilize these features. MulT is particularly effective because it addresses two tasks with overlapping characteristics, such as object boundaries and depth information. By leveraging a common representation, the model ensures that segmentation benefits from accurate spatial relationships, while 3D coordinate prediction improves by utilizing semantic grouping. This complementary design allows the model to achieve high accuracy in both tasks by reinforcing the strengths of shared and task-specific features.

\subsection{Deriving Environmental Feature Set \texorpdfstring{\(F\)}{F}}
After performing 3D coordinate extraction and semantic segmentation, the feature set \( F \) is derived by combining directly obtained values and inferred properties based on the extracted data. We next explain how each component of \( F \) is determined. The 3D coordinates \((x, y, z)\) are directly obtained from the 3D coordinate prediction output which provides the spatial location of each pixel in the object relative to the camera. These coordinates are combined to calculate the object's center or other key points needed for tasks like understanding where objects are or how they are positioned. In this setup, it is assumed that the camera and antenna share the same position. This ensures that the camera information can be directly applied to the wireless communication environment.

Additionally, the azimuth angle $\phi$ and elevation angle $\theta$, which describe the object's relative direction from the camera and thus the antenna, are derived using:
\begin{equation}
\phi = \arctan\left(\frac{x}{z}\right), \quad \theta = \arctan\left(\frac{y}{z}\right).
\end{equation}
These angles serve as critical parameters for representing the object's position in the camera's (and antenna's) field of view. Combining 3D coordinates with azimuth and elevation angles provides a complete view of an object's position and orientation, improving accuracy in spatial analysis and downstream processing. These angles are critical for determining the object's spatial orientation and serve as inputs for further calculations, such as AoA and AoD.
The object size \((w_j, h_j, d_j)\) is computed by analyzing the bounding box of the object in the 3D space. Specifically, the minimum and maximum 3D coordinates for each axis (\(x, y, z\)) are identified from the segmentation mask of the object. The width, height, and depth are then given by:
\begin{equation}
w = x_{\text{max}} - x_{\text{min}}, \quad h = y_{\text{max}} - y_{\text{min}}, \quad d = z_{\text{max}} - z_{\text{min}}.
\end{equation}
This process provides the physical dimensions of the object, which are essential for estimating its impact on signal propagation, such as reflection and diffraction. The material properties \( m_j \) are obtained directly from the semantic segmentation output, where each pixel is classified into predefined material categories. This property captures the reflective or absorptive characteristics of the object, which are crucial for signal modeling. The spatial distance \( r_j \) between the object \( j \) and the camera is computed directly from the depth map, which provides distance measurements per pixel. The distance for object \( j \) is typically calculated as the mean or median depth value across the pixels belonging to the object:
\begin{equation}\label{eq:22}
r_j = \frac{1}{N_j} \sum_{p \in \mathcal{M}_j} d_p,
\end{equation}
where \( \mathcal{M}_j \) represents the segmentation mask of object \( j \), which includes all pixels belonging to the object. \( d_p \) is the depth value of pixel \( p \) within the mask \( \mathcal{M}_j \), and \( N_j \) is the total number of pixels in the mask. (\ref{eq:22}) allows us to compute the object's representative distance \( r_j \) as the average depth value of all its pixels. This ensures an accurate measurement of its spatial distance from the camera.
These parameters describe the directions of the signal arrival and departure relative to the object, which are essential for modeling signal propagation paths. Since we use Carla-generated data, the distances and dimensions are in actual physical units such as meters. Carla provides ground truth values for the 3D environment to ensure that the extracted features \( (w_j, h_j, d_j) \) and \( r_j \) represent the real-world dimensions and distances. The physical accuracy of this data is critical for realistic channel modeling and signal analysis, as it allows the feature set \( F \) to directly reflect the characteristics of the actual urban environment.

\subsection{Structural Causal Model for Channel Estimation}
The environmental features \( F_k \), extracted using the MulT model, represent object distances, dimensions, and material properties in the communication environment at time slot \( k \). To model the relationships between these features and the channel parameters \( X_k \), we employ a Structural Causal Model(SCM), which captures the dependencies between \( F_k \), \( X_k \), and the channel response \( \boldsymbol{H}_k \). To represent the deterministic relationships between the environmental variables \( F_k \), channel parameters \( X_k \), and channel response \( \boldsymbol{H}_k \), we construct a SCM \cite{peters2017elements}. The SCM represents these variables as nodes in a directed acyclic graph (DAG) \( G \), where the edges signify causal dependencies. This structure systematically captures both the direct and indirect effects of environmental factors on the channel parameters and the channel response.

\subsubsection{Structure of the SCM}
The SCM $\mathcal{G} = (V, E)$ is defined with the variable set \( V = \{E_{\alpha,k}, E_{\beta,k}, \dots, E_{\zeta,k}, X_k, \boldsymbol{H}_k\} \), including the environmental variables \( E_k = \{E_{\alpha,k}, E_{\beta,k}, \dots, E_{\zeta,k}\} \), channel parameters \( X_k = \{X_{1,k}, X_{2,k}, X_{3,k}\} \), and the channel response \( \boldsymbol{H}_k \). The edges \( E \) represent causal relationships, specifically:
\begin{equation}\label{eq:24}
E_k \to X_k \to \boldsymbol{H}_k.
\end{equation}
The joint distribution \( p(E_k, X_k, \boldsymbol{H}_k) \) is factorized as:
\begin{equation}\label{eq:25}
p(E_k, X_k, \boldsymbol{H}_k) = p(E_k) \cdot \prod_{\mu=1}^3 p(X_{\mu,k} \mid E_k) \cdot p(\boldsymbol{H}_k \mid X_k),
\end{equation}
where \( p(E_k) \) describes the joint distribution of the environmental variables extracted from observations at time \( k \). The conditional distribution \( p(X_{\mu,k} \mid E_k) \) maps the environmental variables \( E_k \) to the channel parameters \( X_{\mu,k} \), effectively capturing their dependencies. Finally, \( p(\boldsymbol{H}_k \mid X_k) \) defines the physical relationship that determines the channel response \( \boldsymbol{H}_k \) from the channel parameters \( X_k \). Here, the environmental features \( F_k \) serve as a high-level representation that summarizes the detailed environmental variables \( E_k \). While \( E_k \) includes granular data such as object positions, dimensions, and material properties, \( F_k \) extends this information with additional contextual features necessary for causal modeling and channel estimation. The causal structure \eqref{eq:24},\eqref{eq:25} is initially formulated based on channel model in \eqref{eqn:channelmodel}. However, instead of relying solely on predefined equations, VCD refines the causal relationships by leveraging observational data to capture real-world variations. VCD updates only the affected causal components when environmental changes occur rather than retraining the entire model.

\subsubsection{Causal Relationships in SCM}
The environmental features \( E_k = \{E_{\alpha,k}, E_{\beta,k}, \dots, E_{\zeta,k}\} \) causally influence the channel parameters \( X_k = \{X_{1,k}, X_{2,k}, X_{3,k}\} \) through deterministic functions:
\begin{equation}
X_{\mu,k} = g_\mu(E_{\alpha,k}, E_{\beta,k}, \dots, E_{\zeta,k}), \quad \forall \mu \in \{1, 2, 3\}.
\end{equation}
Here, \( X_{1,k}, X_{2,k}, X_{3,k} \) represent the path gain, AoA/AoD, and path distance at time slot \( k \), respectively. These functions capture how material properties (\( E_{\gamma,k} \)), spatial arrangements (\( E_{\delta,k}, E_{\epsilon,k} \)), and relative object positions (\( E_{\alpha,k}, E_{\beta,k}, E_{\chi,k} \)) influence \( X_k \). For instance, \( E_{\gamma,k} \) determines reflectivity, while \( E_{\delta,k}, E_{\epsilon,k} \) affect propagation angles, and \( E_{\alpha,k}, E_{\beta,k}, E_{\chi,k} \) govern distances.

\subsubsection{Learning the SCM}
The causal graph \( G \) is parameterized by an adjacency matrix \( M_G \), which encodes the causal relationships between the environmental features \( F_k = \{E_{a,k}, E_{b,k}, \dots, E_{n,k}\} \) and the channel parameters \( X_k \). Each element of \( M_G \) is defined as:
\begin{equation}
M_G[\alpha, \beta] = 
\begin{cases} 
1, & \text{if } E_{\alpha,k} \text{ causally influences } X_{\beta,k}, \\ 
0, & \text{otherwise}.
\end{cases}
\end{equation}
Here, \( E_{\alpha,k} \) represents a specific environmental feature at time slot \( k \), such as material properties, spatial positions, or object velocities, while \( X_{\beta,k} \) corresponds to a channel parameter. The sparse mechanism shift (SMS) hypothesis ensures that \( M_G \) remains sparse, reflecting the realistic assumption that environmental changes typically influence only a subset of causal mechanisms. For instance, a change in object material properties (\( E_{6,k} \)) may primarily affect path gain (\( X_{1,k} \)) but have minimal impact on AoA or path distance (\( X_{2,k}, X_{3,k} \)). This sparsity enhances the interpretability of the SCM and facilitates efficient adaptation to new scenarios, as only a limited number of affected mechanisms need to be adjusted when the environment changes.

\subsubsection{Causal Inference and Integration with VCD}
Using the learned SCM, causal inference estimates the channel parameters at time \( k \):
\begin{equation}
p(X_k \mid F_k) = \prod_{\beta=1}^3 p(X_{\beta,k} \mid \text{PA}_{\beta,k}),
\end{equation}
where \( \text{PA}_{\beta,k} \) represents the parents of \( X_{\beta,k} \) in \( G \), as determined by the adjacency matrix \( M_G \). Each \( p(X_{\beta,k} \mid \text{PA}_{\beta,k}) \) is modeled as a neural network that incorporates the features of the parent nodes. 

The SCM provides the foundation for the VCD framework. The latent representation \( z_k \), derived from \( F_k \), summarizes the environmental features and encodes \( G \), facilitating efficient causal inference. \( z_k \) captures both the temporal dynamics of the environment and sparse changes due to interventions. The transition dynamics of \( z_k \) are modeled as:
\begin{equation}
p(z_k \mid z_{k-1}, a_{k-1}) = \prod_{\nu=1}^d p_\nu(z_k^\nu \mid M_G^\nu \odot [z_{k-1}, a_{k-1}]),
\end{equation}
where \( M_G^\nu \) determines the causal parents of \( z_k^\nu \) by selecting relevant components from the previous latent state \( z_{k-1} \) and the intervention \( a_{k-1} \). This ensures that only a subset of relevant factors influences each component \(z_k ^ \nu \), maintaining the sparsity and interpretability of the model.

\subsection{VCD for Channel Estimation}
Conventional learning-based channel estimation methods primarily learn statistical relationships between observations and channel variables and consequently have limited ability to adapt to dynamic environments. However, urban THz communications involve rapidly changing factors, such as moving obstacles and structural variations, that significantly impact channel characteristics over time. To address these challenges, we adopt VCD, which simultaneously captures temporal dynamics and causal dependencies for robust channel estimation.  VCD extends SCM by incorporating a time-evolving latent state $z_k$ that models environmental features and their causal effects on channel parameters. Unlike conventional static approaches, VCD dynamically identifies environmental changes, such as the emergence of obstacles or modifications in building structures, and selectively updates only the affected components. This allows the model to maintain high estimation accuracy in new environments without requiring complete retraining. The following subsections detail how VCD models these environmental transitions and effectively learn the channel variables $\boldsymbol{X}_k$ for accurate channel estimation.

\subsubsection{Latent State Transition Modeling}
Channel characteristics in THz communication continuously evolve due to environmental dynamics. Small changes, such as vehicle and pedestrian movement, gradual shifts in reflection surfaces, or variations in diffraction effects at building edges, can significantly impact signal propagation over time. These variations do not occur instantaneously but follow a smooth transition, which means that the channel state at any given moment is strongly influenced by past conditions. Conventional models that treat each observation independently fail to capture this temporal continuity, which leads to abrupt estimation errors and requires frequent retraining in new environments.

To incorporate these evolving dependencies, we introduce a latent state \( z_k \) that evolves over time to ensure that past information is preserved and only relevant components are updated. Unlike static models that discard previous observations, this approach maintains continuity in estimation, which allows the model to track gradual environmental changes rather than reacting only to instantaneous measurements.

The transition of \( z_k \) is modeled as:
\begin{equation}
p(z_k \mid z_{k-1}, a_{k-1}) = \prod_{\nu=1}^d p_\nu(z_k^\nu \mid M_G^\nu \odot [z_{k-1}, a_{k-1}]),
\end{equation}
where \( M_G^\nu \) enforces causally relevant updates to ensure that only affected components are adjusted while preserving prior knowledge.

Each component \( p_\nu \) follows a Gaussian distribution:
\begin{equation}
p_\nu(z_k^\nu \mid h_k^\nu) = \mathcal{N}(\mu_\nu(h_k^\nu), \sigma_\nu^2(h_k^\nu)),
\end{equation}
where \( h_k^\nu \) is a recurrent hidden state that captures temporal dependencies.

By modeling \( z_k \) as a stochastic variable, the system accounts for uncertainties in environmental changes to reduce the need for retraining while improving adaptation.
This enables robust channel estimation even in highly dynamic THz environments, where minor environmental variations can significantly alter signal propagation.

\subsubsection{Sparse Mechanism Shift Hypothesis}
While the latent state transition model enables the tracking of environmental variations over time, it implicitly assumes that all components of the latent state \( z_k \) evolve continuously. However, in real-world THz environments, changes do not affect all aspects of the channel equally. Instead, environmental shifts such as the movement of obstacles or structural modifications tend to influence only a subset of the underlying mechanisms, while others remain invariant. 

To capture this sparsity in environmental shifts, we adopt the SMS hypothesis, which states that only specific latent components are affected by external changes while the remaining components retain their prior transition dynamics. This assumption reflects the reality that a new obstacle may alter reflection paths but leave diffraction effects unchanged, or a moving vehicle may introduce blockage without modifying large-scale fading characteristics.
To formalize this, we introduce an intervention mask \( R_I \), which selectively updates only the affected components while preserving the rest. The transition dynamics in a new environment are expressed as:
\begin{equation}
p_{\text{new}}(z_k \mid z_{k-1}, a_{k-1}) = \prod_{\nu=1}^d \big[ (1 - R_{I_\nu}) p_\nu^{(0)}(z_k^\nu) + R_{I_\nu} p_\nu^{(k)}(z_k^\nu) \big],
\end{equation}
where \( R_{I_\nu} \) is a binary mask indicating whether the \( \nu \)-th latent dimension \( z_k^\nu \) is influenced by an intervention. \( p_\nu^{(0)}(z_k^\nu) \) represents the shared observational model that remains unchanged, while \( p_\nu^{(k)}(z_k^\nu) \) accounts for the modified transition dynamics under the new environmental conditions.
By leveraging this sparse adaptation mechanism, the model ensures computational efficiency and stability in dynamic environments. Instead of retraining the entire system when changes occur, only the necessary components are adjusted, allowing the model to generalize effectively across different THz scenarios while maintaining high estimation accuracy.
\begin{algorithm}[ht]
\caption{Vision and Causal Learning-Based Channel Estimation}
\label{alg:channel_estimation}
\begin{algorithmic}[1]
\State \textbf{Input:} RGB-D images.
\State \textbf{Output:} Channel response \( \boldsymbol{H}_k \).

\State Extract environmental features \( F_k \) from RGB-D images:
\[
F_k = \{(x_j, y_j, z_j), r_j, (\text{AoA}_j, \text{AoD}_j), (w_j, h_j, d_j), m_j, v_j\}.
\]

\State Construct causal relationships:
\[
F_k \to X_k, \quad X_k \to \boldsymbol{H}_k.
\]

\State Learn the causal adjacency matrix \( M_G \) using the Sparse Mechanism Shift (SMS) hypothesis.

\State Model the latent state transitions:
\[
p(z_k \mid z_{k-1}, a_{k-1}) = \prod_{i=1}^d p_i(z_k^i \mid M_G^i \odot [z_{k-1}, a_{k-1}]).
\]

\State Optimize the Evidence Lower Bound (ELBO) to learn latent representations \( z_k \):
\[
\text{ELBO} = \mathbb{E}_{q_\phi(z_k \mid o_k)}[\log p_\theta(o_k \mid z_k)] - \text{KL}[q_\phi \| p].
\]

\State Decode the latent state \( z_k \) to estimate channel variables \( X_k \):
\[
X_k = f_{\text{channel}}(z_k),
\]
where \( X_{1,k} = \gamma_k^{(l)} \), \( X_{2,k} = \{\phi_{r,k}^{(l)}, \phi_{t,k}^{(l)}\} \), and \( X_{3,k} = d_k^{(l)} \).

\State Compute the channel response \( \boldsymbol{H}_k \) using the channel model:
\[
\boldsymbol{H}_k = \sum_{l=0}^L \gamma_k^{(l)} \eta(d_k^{(l)}) \boldsymbol{a}_r(\phi_{r,k}^{(l)}) \boldsymbol{a}_t^H(\phi_{t,k}^{(l)}).
\]
\end{algorithmic}
\end{algorithm}

\subsubsection{Learning Objectives and Evidence Lower Bound (ELBO) Optimization}

The SMS hypothesis suggests that only specific components of the latent state \( z_k \) are influenced by environmental changes, while the remaining components retain their previous dynamics. To ensure that the learning process adheres to this principle, we optimize the evidence lower bound (ELBO), which encourages the model to capture meaningful latent representations while preserving the underlying causal structure.

The ELBO objective is defined as:
\begin{equation}\label{eqn:ELBO}
\begin{aligned}
\text{ELBO} = & \sum_{k=0}^T \mathbb{E}_{q_\phi(z_k \mid o_k)} \big[ \log p_\theta(o_k \mid z_k) \big] \\
& - \mathbb{E}_{q_\phi(z_{k-1} \mid o_{k-1})} \big[ \text{KL}\big( q_\phi(z_k \mid o_k) \| p(z_k \mid z_{k-1}, a_{k-1}) \big) \big].
\end{aligned}
\end{equation}
The first term, \( \log p_\theta(o_k \mid z_k) \), represents the reconstruction loss, ensuring that the learned latent state \( z_k \) effectively reconstructs the observed data \( o_k \). The second term, the KL divergence \( \text{KL}[ q_\phi \| p ] \), regulates the approximate posterior \( q_\phi(z_k \mid o_k) \) to match the prior \( p(z_k \mid z_{k-1}, a_{k-1}) \) and preserve the causal structure inferred from the SCM.

By optimizing the ELBO, the model learns a structured latent representation that aligns with both observed features and the sparsity constraints imposed by the SMS hypothesis. This ensures that only the necessary components are updated in response to environmental shifts, which reduces unnecessary parameter adjustments and improves generalization to new conditions without requiring retraining.

\subsubsection{Channel Variable Estimation via Variational Autoencoder}
The latent state \( z_k \), obtained by modeling the latent state transition, encodes the temporal evolution of the environment while preserving causal relationships among the factors that affect the channel. However, \( z_k \) itself does not directly correspond to the physical channel parameters. To accurately estimate the channel response, it is necessary to transform \( z_k \) into meaningful variables such as the gain of the path \( X_{1,k} = \gamma_k^{(l)} \), the angles of arrival and departure \( X_{2,k} = \{\phi_{r,k}^{(l)}, \phi_{t,k}^{(l)}\} \), and the distance from the path \( X_{3,k} = d_k^{(l)} \).

To achieve this, a variational autoencoder (VAE) is employed. The VAE introduces a structured latent space that captures the underlying properties of the environment while incorporating uncertainty. Unlike conventional regression models, which rely on deterministic mappings, the VAE ensures that channel variable estimations remain robust even under unobserved conditions by leveraging probabilistic inference. The encoder network approximates the posterior distribution of the latent state:

\begin{equation}
q_\phi(z_k \mid o_k) = \mathcal{N}(z_k; \mu_\phi(o_k), \sigma^2_\phi(o_k)),
\end{equation}

where \( \mu_\phi(o_k) \) and \( \sigma^2_\phi(o_k) \) are the mean and variance parameterized by observations \( o_k \). Instead of directly mapping observations to a deterministic latent state, the model samples from the learned distribution.

\begin{equation}
z_k = \mu_\phi(o_k) + \sigma_\phi(o_k) \odot \epsilon, \quad \epsilon \sim \mathcal{N}(0, I).
\end{equation}

Given the inferred latent state \( z_k \), the decoder reconstructs the channel variables by modeling the conditional likelihood:

\begin{equation}
p_\theta(X_k \mid z_k) = \mathcal{N}(X_k; \mu_\theta(z_k), \sigma^2_\theta(z_k)).
\end{equation}

To ensure accurate reconstruction of the channel variables, the VAE is trained to minimize the mean squared error (MSE) between the predicted and ground truth values:

\begin{equation}
\text{MSE}_X = \frac{1}{N} \sum_{i=1}^{N} \sum_{j=1}^{J} \| \hat{X}_{i,j} - X_{i,j} \|^2.
\end{equation}

Since the final channel response \( H_k \) relies on the accuracy of the estimated $X_k$, we also measure $\text{MSE}_H$ to evaluate the end-to-end performance of our approach.
\begin{equation}
\text{MSE}_H = \frac{1}{N} \sum_{i=1}^{N} \| \hat{H}_i - H_i \|^2.
\end{equation}

Minimizing \( \text{MSE}_X \) leads to lower \( \text{MSE}_H \), ensuring that accurate estimation of channel parameters contributes to better channel prediction.

Unlike conventional VAE applications, where the decoder reconstructs raw data, the decoding process here must ensure that the estimated channel parameters adhere to the physical constraints of the propagation environment. To achieve this, the decoder is designed to preserve structured relationships between \( z_k \) and \( X_k \), ensuring that the learned mapping maintains consistency with fundamental wireless propagation. The estimated channel variables are expressed as follows:

\begin{equation}
X_{1,k} = f_{\gamma}(z_k) = \sum_{i=1}^{M} w_{\gamma,i} g_{\gamma,i}(z_k),
\end{equation}

\begin{equation}
X_{2,k} = f_{\phi}(z_k) = \sum_{i=1}^{M} w_{\phi,i} g_{\phi,i}(z_k),
\end{equation}

\begin{equation}
X_{3,k} = f_d(z_k) = \sum_{i=1}^{M} w_{d,i} g_{d,i}(z_k),
\end{equation}

where \( g_{\gamma,i}(z_k) \), \( g_{\phi,i}(z_k) \), and \( g_{d,i}(z_k) \) represent nonlinear transformations that reflect propagation characteristics such as reflection, diffraction, and scattering. The weight parameters \( w_{\gamma,i} \), \( w_{\phi,i} \), and \( w_{d,i} \) are optimized to ensure that the transformations remain consistent with the physical properties of the channel. This formulation allows the decoder to infer channel parameters in a structured manner while preserving their physical meaning.

Unlike the generic VCD in~\cite{lei2022variational}, where latent states and decoders are defined in a task-agnostic manner, our framework explicitly adapts the VCD structure to THz channel estimation. First, the latent state is designed to directly encode physically meaningful channel parameters (path gain, AoA/AoD, distance) rather than abstract features. Second, the transition function \(p(z_k \mid z_{k-1}, a_{k-1})\) is constrained by the Sparse Mechanism Shift (SMS) hypothesis, ensuring that only a subset of mechanisms is updated under environmental changes, in contrast to the joint evolution assumed in~\cite{lei2022variational}. Finally, the decoder follows a hierarchical causal order (path gain \(\rightarrow\) angles \(\rightarrow\) distance), preserving propagation consistency and physical interpretability.

To further enhance interpretability and maintain causal dependencies among channel variables, the decoding process follows a hierarchical structure.

\begin{equation}
\begin{aligned}
p(X_k \mid z_k) &= p(X_{1,k} \mid z_k) \\
&\quad \times p(X_{2,k} \mid X_{1,k}, z_k) \\
&\quad \times p(X_{3,k} \mid X_{2,k}, X_{1,k}, z_k).
\end{aligned}
\end{equation}

This hierarchical formulation ensures that the estimation follows a physically valid order. The path gain \( X_{1,k} \), which depends on reflection and diffraction effects, is first estimated. The AoA and AoD \( X_{2,k} \) are then conditioned on \( X_{1,k} \) to ensure multipath consistency. The path distance \( X_{3,k} \) is determined based on both the path gain \( X_{1,k} \) and the angles of arrival and departure \( X_{2,k} \). Since signals travel along different paths due to reflections and diffractions, the estimated distance accounts for these effects rather than assuming a direct line between the transmitter and receiver.
By explicitly enforcing this dependency structure, the model guarantees that the estimated parameters remain physically plausible rather than treating them as independent variables.

An additional advantage of this structured inference approach is its ability to adapt to environmental variations without requiring complete retraining. This is achieved through the Sparse Mechanism Shift Hypothesis (SMS), which states that only a subset of mechanisms governing channel evolution change over time. Instead of modifying the entire model when environmental conditions shift, the VAE selectively updates only the affected latent dimensions. This is formulated as follows:

\begin{equation}
p_{\text{new}}(z_k \mid z_{k-1}, a_{k-1}) = \prod_{\nu=1}^d \big[ (1 - R_{I_\nu}) p_\nu^{(0)}(z_k^\nu) + R_{I_\nu} p_\nu^{(k)}(z_k^\nu) \big],
\end{equation}

where \( R_{I_\nu} \) is a binary mask that determines whether the \( \nu \)-th latent dimension \( z_k^\nu \) undergoes an intervention due to environmental changes. By applying this selective adaptation, the model ensures that variables unaffected by the shift retain their prior distributions while only the necessary components are adjusted to reflect new propagation conditions. This enables the model to efficiently generalize to unseen environments while minimizing computational overhead.

Through this structured inference process, the VAE-based channel estimation framework effectively maps latent representations to physical channel parameters while maintaining causal dependencies and ensuring robustness under real-world THz propagation conditions.
When adapting to a new environment, the model identifies the set of intervened mechanisms \( R_I \) and updates only the corresponding parameters:
\begin{equation}
\mathcal{L}_{\text{adapt}} = \mathbb{E}_{G, I} \Big[ \text{ELBO}(o_{0:k}^{\text{new}}, a_{0:k}^{\text{new}}; \theta, \phi, G, I) - \lambda_I \|I\| \Big].
\end{equation}
This approach takes advantage of the sparsity of \( R_I \) to efficiently adapt the model while reusing shared mechanisms.7{At deployment in unseen environments, the intervention mask \(R_I\) is not assumed known. It is inferred online from a short window of new observations. When the data is consistent with the shared components learned during training, the mask remains inactive, and when significant deviations are detected, only environment-specific mechanisms are sparsely updated. This lightweight adaptation realizes SMS in practice without retraining the entire model. Finally, the channel response \(\boldsymbol{H}_k\) is computed using equation (\ref{eqn:approximate}). This approach integrates the GCM structure, VCD dynamics, and physical models to provide a robust and interpretable solution for channel estimation.

\subsection{Complexity Analysis}
In this section, we briefly analyze the computational complexity of the proposed pipeline in comparison with standard baselines introduced in section V. CNN estimators scale as $\mathcal{O}(HW\,C_{\text{in}}C_{\text{out}}k^2)$ and run efficiently on GPUs~\cite{8114708}. LSTMs require $\mathcal{O}(T d^2)$ and suffer from sequential updates. Transformers/MAE incur $\mathcal{O}(N^2 d)$ for full attention, reducible to $\mathcal{O}(N w d)$ with windowed attention~\cite{HeCVPR2022MAE}. GNNs scale roughly as $\mathcal{O}(L|E|d)$ per layer~\cite{9046288}. Matrix completion relies on iterative singular value decomposition(SVD) with $\mathcal{O}(I r M N_m)$ (or higher) per iteration~\cite{8466658}, typically heavier than a single transformer pass. Our MulT encoder follows transformer-like scaling with reduced tokens (patching, windows), and the VCD head adds only $\mathcal{O}(E_c d)$; overall, the cost is comparable to a lightweight transformer plus a small graph update. Here, $H{\times}W$ is input size, $C_{\text{in}}/C_{\text{out}}$ input/output channels, $k$ kernel size, $T$ sequence length, $d$ embedding size, $N$ tokens, $w$ attention window, $L$ layers, $|E|$ graph edges, $M{\times}N_m$ MC matrix size, $r$ rank, $I$ iterations, and $E_c$ causal edges. Training is offline and compute-intensive, but deployment-time inference is parallelizable and completes in milliseconds on modern accelerators~\cite{8322184}. In addition, existing compression (pruning, quantization, distillation) further reduces latency and memory~\cite{8253600}. Furthermore, the framework runs on hardware already assumed for DL-based CSI feedback and hybrid beamforming.

\section{Simulation Results and Analysis}
\begin{figure*}
    \centering
    \includegraphics[width=0.7\textwidth]{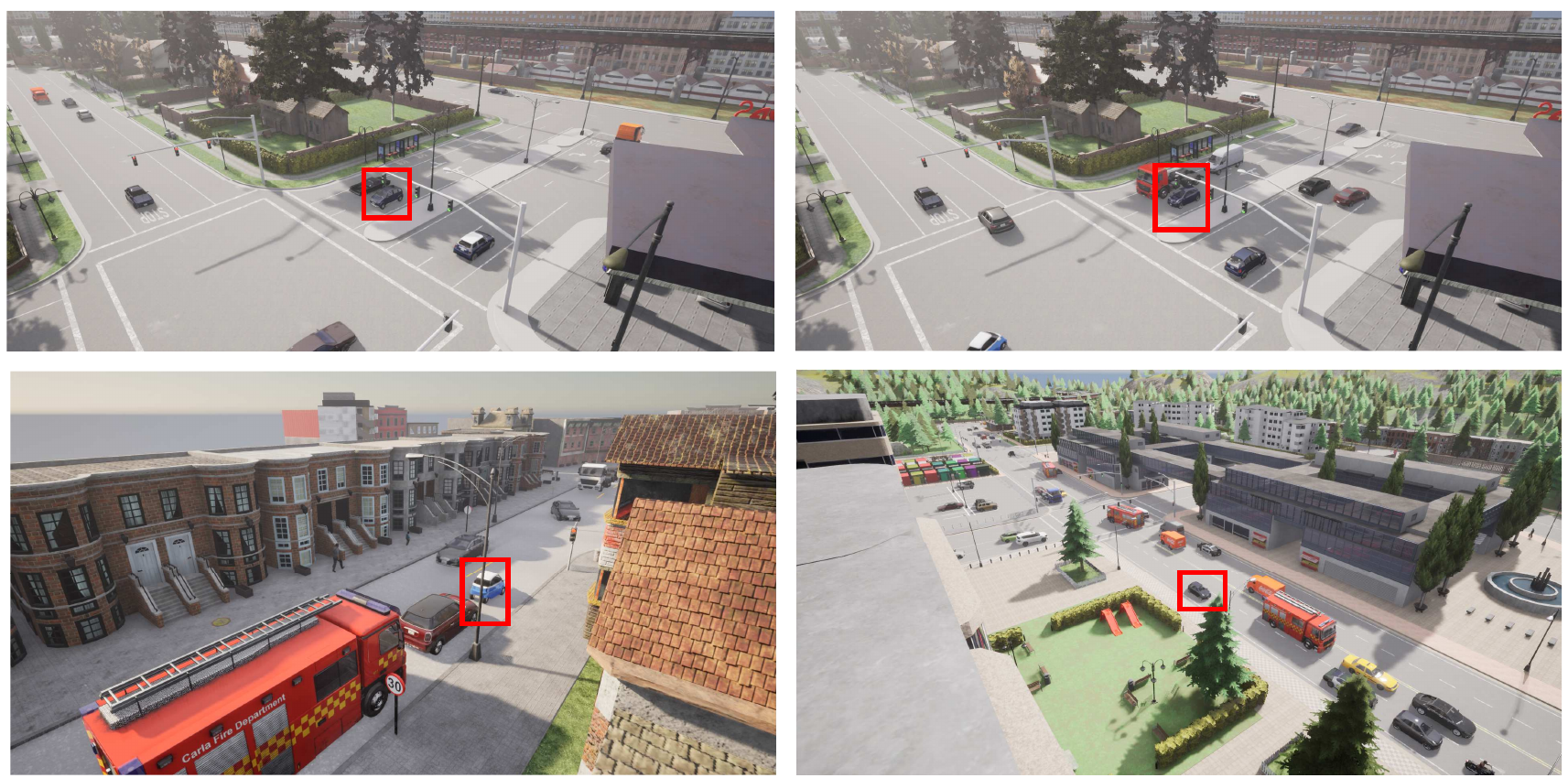}
    \caption{Environmental views from the BS to target vehicles in different scenarios(The target vehicle is in the red box).}
    \label{fig:env}
\end{figure*}

\subsection{Simulation Setup}
We conduct extensive simulations to evaluate the proposed approach in realistic urban THz communication environments. Our goal is to assess how well the model generalizes dynamic urban scenarios with varying environmental conditions. To achieve this, we integrate RGB-D visual data with channel estimation by simulating complex urban landscapes where moving vehicles, buildings, and other obstacles influence THz signal propagation. 

Using MATLAB Ray Tracing and CARLA-generated environments, we model dynamic layouts with multiple vehicles and obstacles to analyze the impact of reflections, scattering, and path loss in different urban settings. The carrier frequency is set to \(0.1\) THz to represent the operating frequency of the communication system. The channel model considers up to five distinct signal paths per UE while limiting reflections to a maximum of one to maintain computational efficiency. The velocity of the UEs ranges from 10 to 50 km/h. The channel variables extracted from this CARLA-based ray-tracing setup are used as ground truth in our experiments. Note that the simulations in this paper constitute an illustrative setup focused on a specific scenario and channel model, intended to demonstrate how the proposed framework operates and where it applies. This setup does not limit the generality of the framework. The proposed framework is model agnostic and can be applied to uncover causal relations between environmental or other factors and channel-model variables, whether they are derived from mathematical formulations, probabilistic models, or measurement-based data. This design enables a deeper understanding of the causes of channel variations and provides a principled basis for channel estimation and prediction.

Urban structures such as buildings, trees, and vehicles are incorporated with well-defined material properties to ensure accurate modeling of signal interactions. Concrete is used for buildings, vegetation for trees, and metal for vehicles. In practice, we obtain scene semantics from CARLA~[36], map the semantic labels (represented by colors) to material categories such as concrete, metal, or vegetation, and assign these materials to object surfaces in the MATLAB ray-tracing engine. The associated propagation parameters are then taken from the MATLAB Ray Tracing Toolbox material library, which allows surfaces to be configured with specific material-related variables. We adopt these settings as reasonable defaults and evaluate the robustness of our estimator to this modeling choice. These materials influence reflection, scattering, and absorption, which are critical for THz signal propagation. Table~I provides a summary of the key parameters. 

\begin{table}[h]
\centering
\caption{Simulation Parameters.}
\setlength{\tabcolsep}{20pt} 
\renewcommand{\arraystretch}{1.5} 
\begin{tabular}{|c||c|}
\hline
\textbf{Parameter}           & \textbf{Value}          \\ \hline
Carrier Frequency            & 0.1 THz                 \\ \hline
Number of Symbols            & 24 KHz                  \\ \hline
Subcarrier Spacing           & 2 GHz                   \\ \hline
Number of Subcarriers        & 240                     \\ \hline
Velocity of UEs (Vehicles)   & \{10, 50\}              \\ \hline
Max Number of Reflections    & 1                       \\ \hline
Max Number of Paths          & 5                       \\ \hline
Building Materials           & Concrete                \\ \hline
Tree Materials               & Vegetation              \\ \hline
Car Materials                & Metal                   \\ \hline
\end{tabular}
\end{table}

The urban environment used in the simulation is derived from Carla\cite{Dosovitskiy17} 3D simulation data, including detailed coordinates of the transmitter, receiver, and various environmental objects. Additionally, Carla's RGB-D camera sensor is employed to capture visual data of the environment, and the camera parameters such as field of view (FOV), focal length, and resolution are configured to align with the simulation requirements. The FOV is set to 90 degrees to capture a broad perspective, and the focal length is adjusted to ensure accurate depth perception. The extracted 3D environment data and camera parameters are converted into a suitable format for MATLAB, where they are used to create the simulation environment. 
We use MATLAB's Ray Tracing to model the propagation of THz signals within the simulated urban environment. The toolbox employs the shoot-and-bounce ray tracing method to capture the multi-path effects such as reflection, diffraction, and scattering. The transmitter is placed at a height of 10 meters, representing a base station, while the receiver is positioned at varying locations to simulate different UE positions.
The simulation parameters include the frequency of the THz signals, antenna characteristics, and the material properties of the environmental objects. The ray tracing simulations produce detailed data on the signal paths, path loss, and received signal strength at each UE position, which are then used to analyze the wireless channel characteristics. By capturing and modeling these complex interactions, we can accurately evaluate the performance of the proposed channel estimation method in realistic urban conditions. For the baseline, we use the following conventional methods without a causality mechanism to assess the channel estimation performance in different environments.

\begin{itemize}

    \item \emph{Matrix Completion (MC)}:
    A classical pilot-only baseline that reconstructs the time–frequency channel grid via nuclear-norm minimization\cite{8466658} using the observed pilot entries. It does not use any scene/geometry information (e.g., BS/UE poses, AoA/AoD, object maps, or material properties) and typically degrades under strong multipath or mobility.
    
    \item \emph{CNN-based Channel Estimator, 32 Pilots}: 
    This method applies a CNN to resource grid images, where channel data is represented in image form. The CNN extracts features from these images to directly estimate channel parameters, relying on the data representation rather than explicitly modeling spatial dependencies or parameter interactions.

    \item \emph{LSTM-based Channel Estimator, 32 Pilots}: 
    We employ LSTM to estimate temporal variations in channel parameters, as it effectively captures time dependencies in the channel state. However, it does not explicitly account for spatial relationships or environmental factors, which are often critical for accurate estimation in complex scenarios.

    \item \emph{Masked autoencoder(MAE, 24 Pilots)}\cite{10137638}: 
    We use a pilot-based channel estimation method based on an MAE. This method uses 24 pilot symbols as reference points. This method operates based on the pilot symbols only and does not take into account any spatial or environmental features but relies purely on the pilot data for channel estimation.

    \item \emph{MAE (32 Pilots)}\cite{10137638}: 
    This method extends the previous MAE-based approach by using 32 pilot symbols instead of 24. Like the 24-pilot version, it does not consider the spatial environment and focuses solely on pilot data to estimate the channel parameters.

   \item \emph{Multi-task + Graph Neural Networks(GNN)}: 
   The Multi-task + GNN method estimates channel parameters by modeling the environment as a graph, with objects as nodes and spatial relations as edges. It learns spatial dependencies that influence signal propagation and improves estimation accuracy. Unlike physical models, it captures data-driven correlations within the spatial structure but does not model causal relationships between environment and channel variations.
\end{itemize}

The experiment consists of training and testing across different environments. 

\begin{itemize}
    \item \emph{Scenario 1}: The proposed and all baseline models are trained on the dataset from this scenario (Fig.~3, top-left).
    \item \emph{Scenario 2}: A variation of Scenario 1 with changes in the number of users and their movement speeds, used to evaluate model performance in a slightly modified environment (Fig.~3, top-right).
    \item \emph{Scenario 3}: A completely unseen environment for both the baseline models and the proposed algorithm (Fig.~3, bottom-left).
    \item \emph{Scenario 4}: Another unseen environment for both the baseline models and the proposed algorithm (Fig.~3, bottom-right).
\end{itemize}

\subsubsection{Intervention-Based Analysis}
This experiment examines the effect of specific environmental factors by actively modifying them while keeping all other conditions constant in the same scenario. We intervene on two adjustable key variables:

\begin{itemize}
    \item \textbf{Number of Paths} : Limited to $L = \{1, 2, 3, 4, 5\}$ to analyze the effect of multipath components.
    \item \textbf{Vehicle Speed} : Set to $v = \{10, 20, 30, 40, 50\}$ km/h to observe the impact of mobility.
\end{itemize}

The channel estimation performance of baselines and the proposed method is compared using $\text{MSE}_H$.

\subsubsection{Counterfactual Analysis}
Unlike intervention, counterfactual analysis explores hypothetical scenarios: \textit{"What if the environment had been different?"} We take an observed scenario (e.g., scenario 1) and infer how the channel estimation would have changed under a different environment (e.g., Scenario 2,3 and 4).

\begin{itemize}
    \item \textbf{Baseline Condition}: The estimated channel response in Scenario 1, where the model was trained, with the number of paths set to 5 and the vehicle speed set to 50 km/h.
    \item \textbf{Counterfactual Condition}: The estimated response in different scenarios (scenario 2, 3, and, 4) while keeping the number of paths (5) and vehicle speed (50 km/h) unchanged, allowing us to analyze the impact of environmental variations on channel estimation.
\end{itemize}

This setup tests the models' ability to generalize by training them on one environment and evaluating their performance in a modified version of the same environment and entirely new environments. Fig. 3 shows environmental views from the BS for each scenario. Furthermore, in the following results, MSE specifically refers to $MSE_h$.

\subsection{Simulation Results and Analysis}
\begin{figure}[ht]
	\centering
	\captionsetup{justification=centering}
	\includegraphics[width=10.0cm, height=6.8cm]{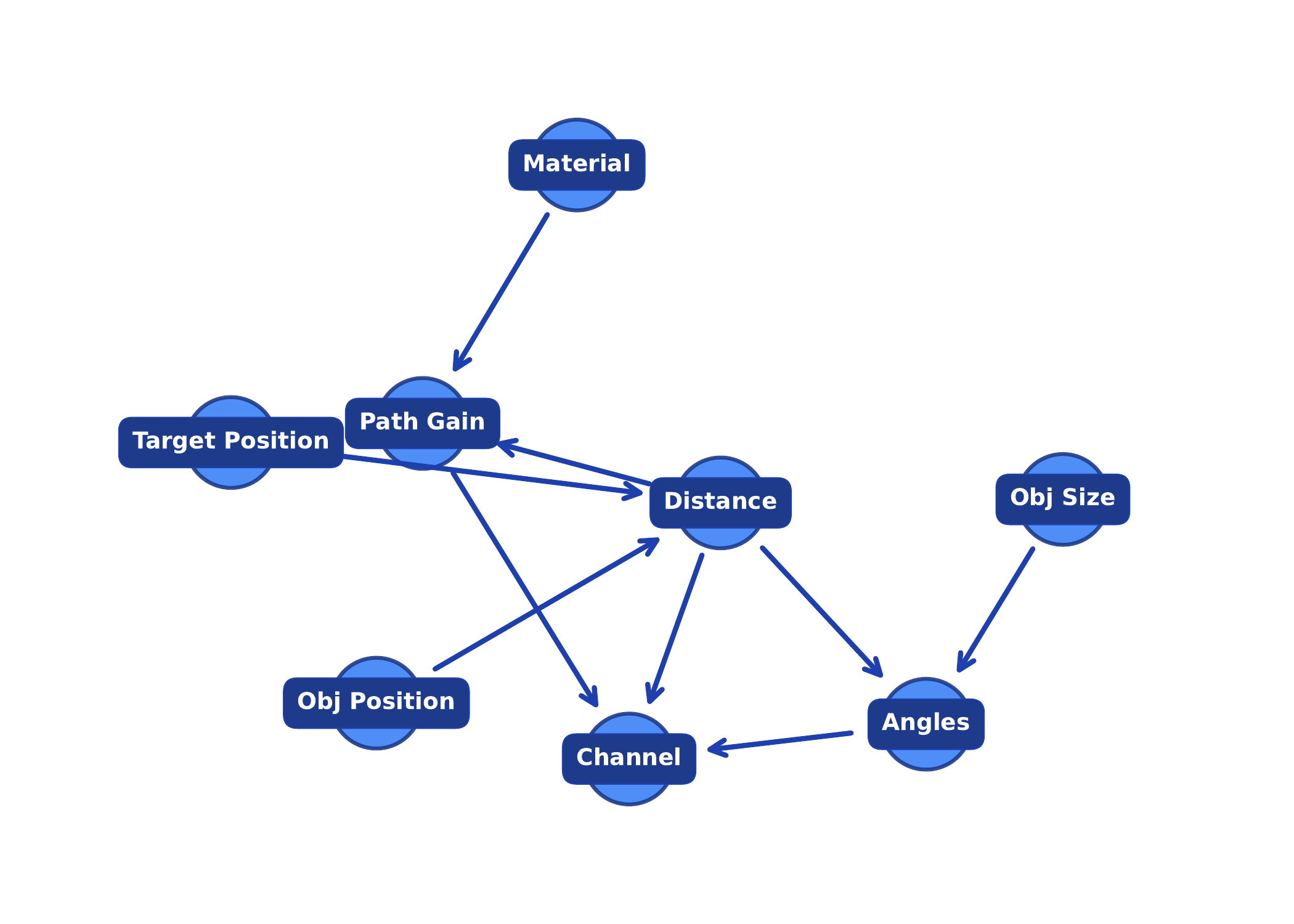}
	\caption{Example of DAG for the channel estimation.}
	\label{fig:DAG}
\end{figure}

\begin{figure*}[t!]
    \centering
    \subfloat[MSE comparison with increasing number of paths (SNR=25dB).]{
        \includegraphics[width=0.38\textwidth]{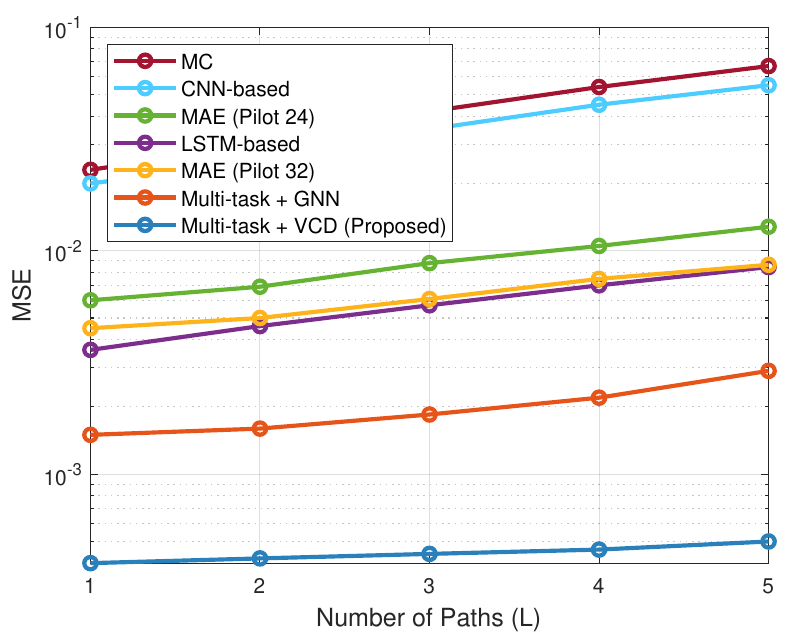}
        \label{fig:intervention_paths}
    }
    \hfil
    \subfloat[MSE comparison with increasing vehicle speed (SNR=25 dB).]{
        \includegraphics[width=0.38\textwidth]{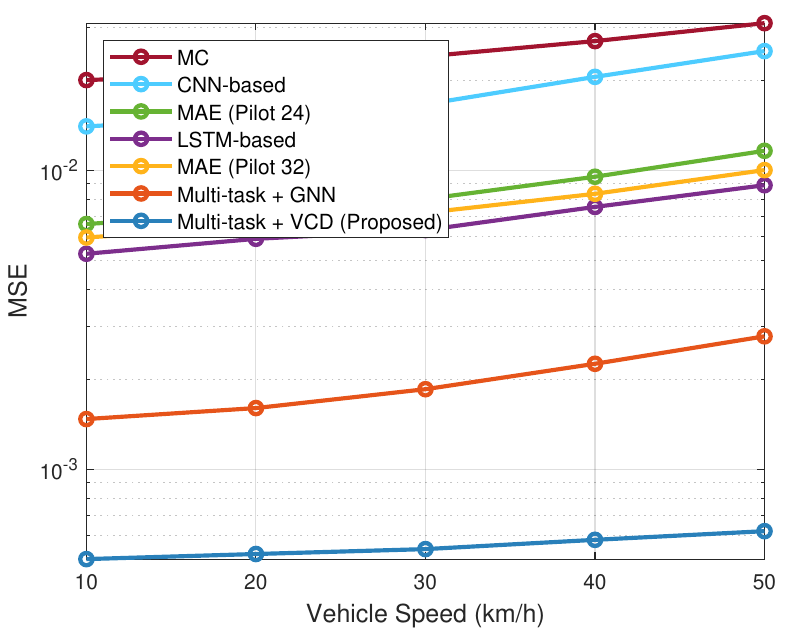}
        \label{fig:intervention_speed}
    }
    \caption{Channel estimation performance comparison across different environmental features in the same scenario (scenario 1).}
    \label{fig:intervention_scenario1}
\end{figure*}

\begin{figure*}[t!]
    \centering
    \subfloat[Scenario 1.]{
        \includegraphics[width=0.38\textwidth]{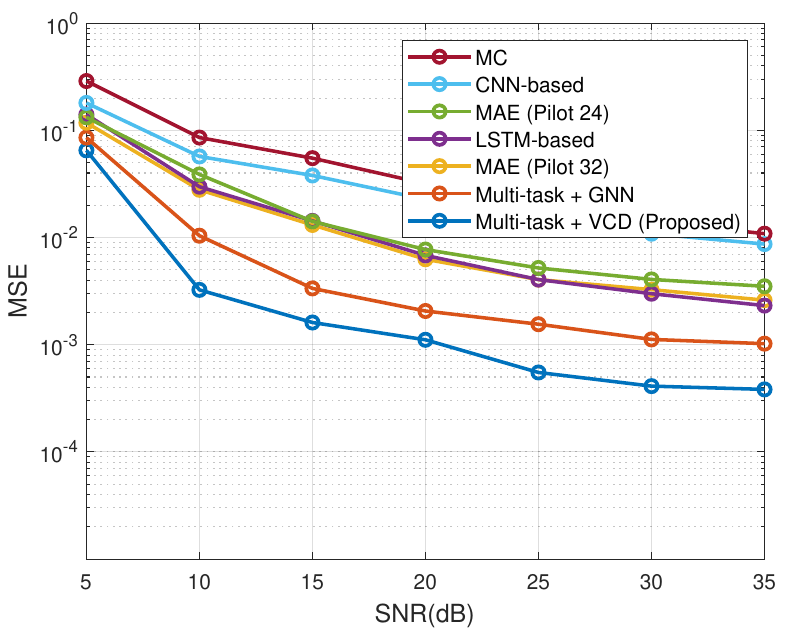}
        \label{fig:scenario1}
    }
    \hfil
    \subfloat[Scenario 2.]{
        \includegraphics[width=0.38\textwidth]{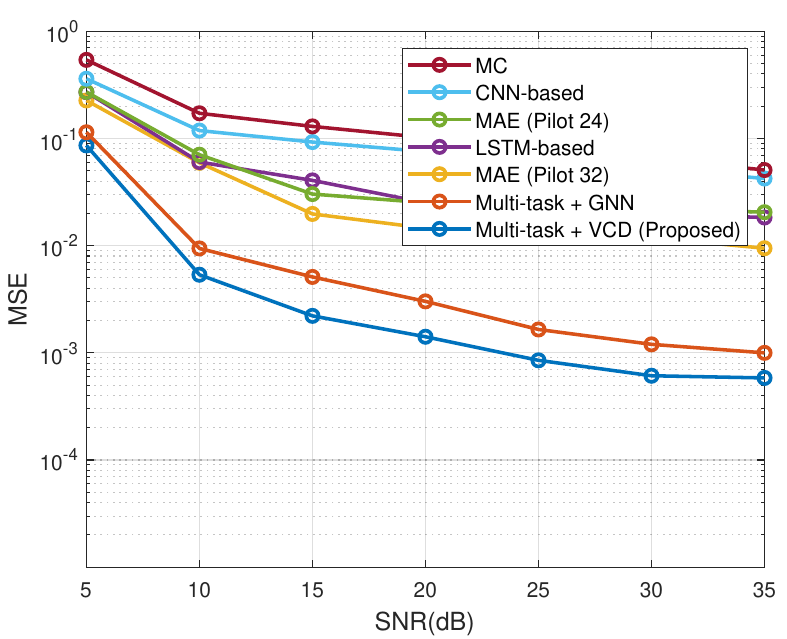}
        \label{fig:scenario2}
    }

    \vspace{0.15in}

    \subfloat[Scenario 3.]{
        \includegraphics[width=0.38\textwidth]{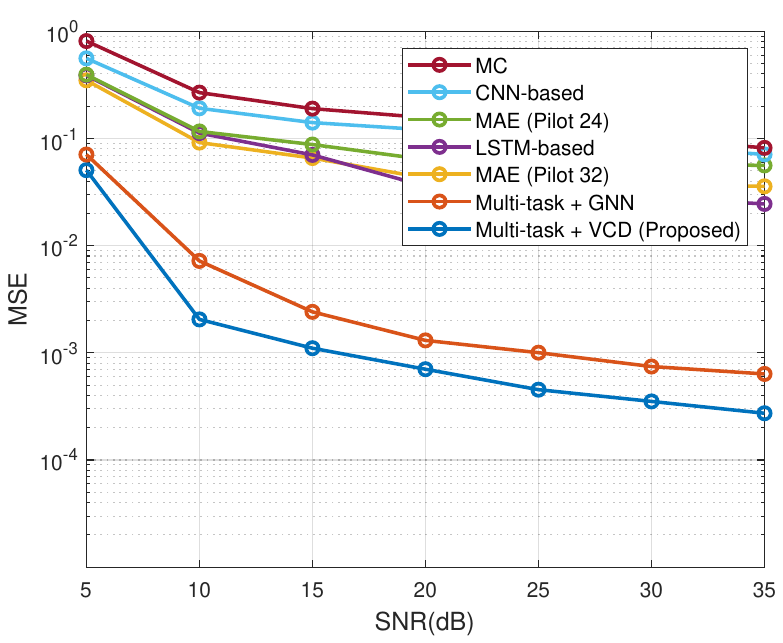}
        \label{fig:scenario3}
    }
    \hfil
    \subfloat[Scenario 4.]{
        \includegraphics[width=0.38\textwidth]{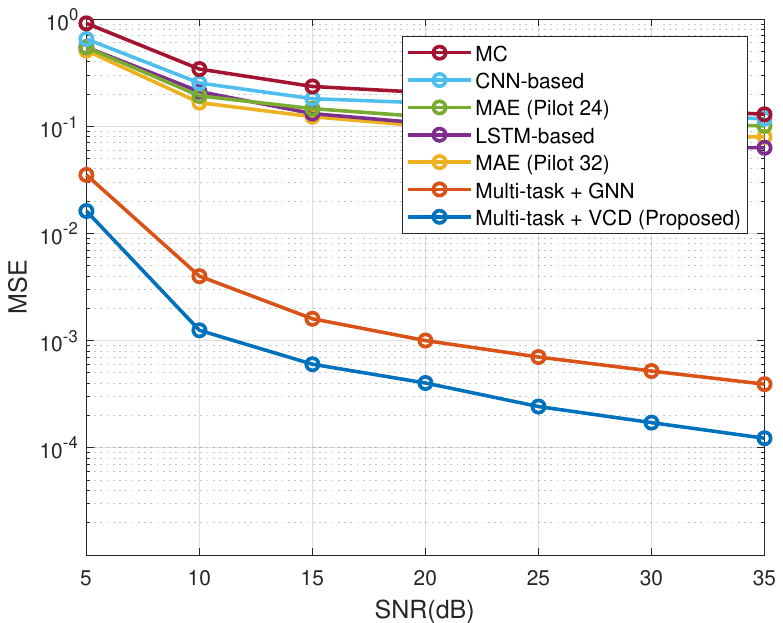}
        \label{fig:scenario4}
    }

    \caption{Channel estimation performance comparison across different scenarios.}
    \label{fig:scenario_comparison}
\end{figure*}

Fig. 4 illustrates the causal relationships learned by the VCD in scenario 1, revealing interactions between environmental and channel variables overlooked by conventional models. While path gain is usually modeled as a monotonic function of distance, the causal graph shows it also depends on angular parameters and environmental factors such as material and object size. These variables affect path gain both directly and indirectly by altering the AoA/AoD distribution, leading to different attenuation patterns even at the same distance. The results further indicate that moving objects cause time-varying changes in AoA/AoD and path gain, demonstrating the VCD’s ability to adapt to dynamic environments in contrast to static conventional models. Validation of these causal dependencies through interventions and counterfactuals will be presented later.

Fig. 5(a) shows the MSE degradation as the number of paths increases at SNR = 25dB. The proposed Multi-task + VCD model acheives the least degradation, while other methods suffer significantly. CNN-based approaches experience the highest degradation, with an increase of approximately 7.3× compared to VCD. The GNN-based method remains more stable than other baselines, with an MSE degradation of 86.6\%. For the MC, MSE also grows as the number of paths increases; because energy spreads over more angles and delays and fine patterns (per-path strength, phase, and cluster width) change, the low-rank fill used by MC misses these angle/delay details, so its degradation is among the largest after CNN. It should be noted that the slight increase in MSE for the proposed method does not indicate a lack of generalization but rather reflects the increased complexity of channel estimation in more challenging environments.

Fig. 5(b) illustrates the MSE degradation with increasing vehicle speed ($v$). The proposed method remains robust, with only a 27.3\% increase in MSE at the highest speed, whereas CNN-based methods exhibit the most significant degradation, exceeding 180\%. For the MC, higher speed makes pilot hints stale and the channel change quickly because MC does not track motion, its error increases faster than GNN/VCD under mobility. Compared to CNN, the proposed VCD model maintains stability, experiencing up to 6.6× lower degradation and ensuring reliable channel estimation even in dynamic environments. The observed minor increase in MSE for the proposed method is a natural consequence of the growing estimation difficulty rather than an indication of degraded generalization ability.


\begin{figure*}
    \centering
    \subfloat[Relationship between channel variables and environmental features in scenario 1.]{
        \includegraphics[width=0.23\linewidth]{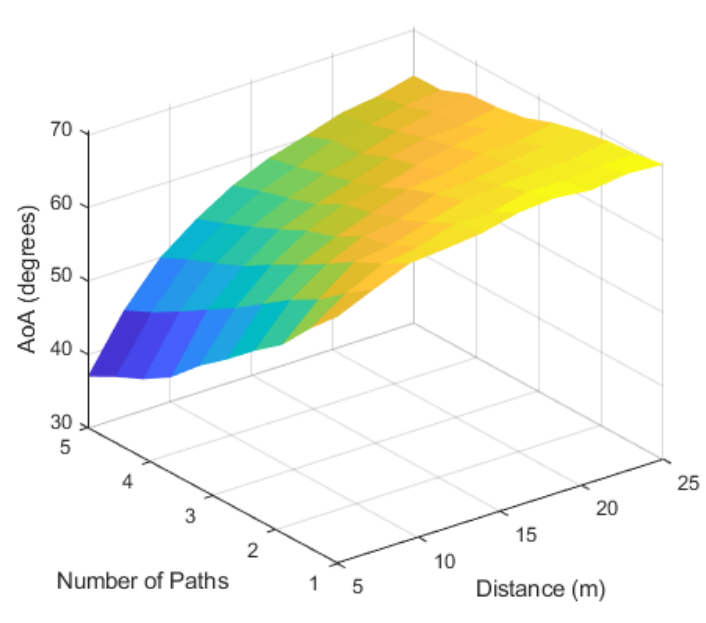}
        \includegraphics[width=0.23\linewidth]{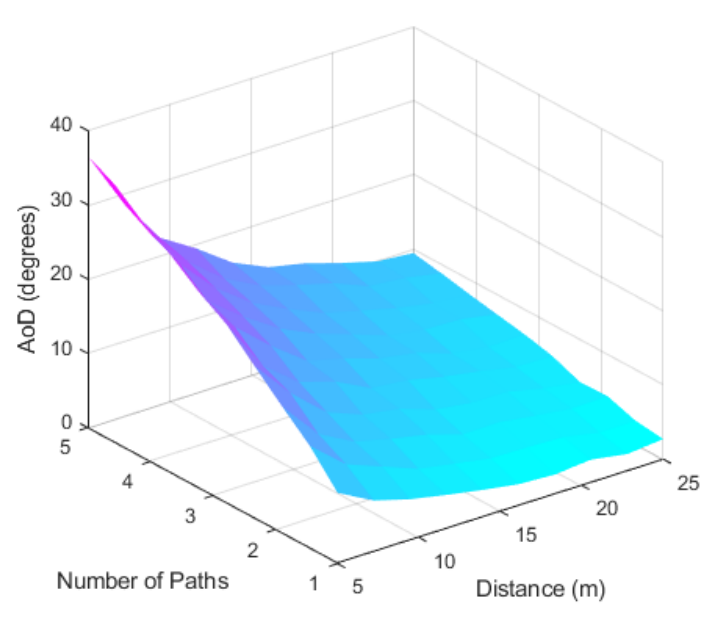}
        \includegraphics[width=0.23\linewidth]{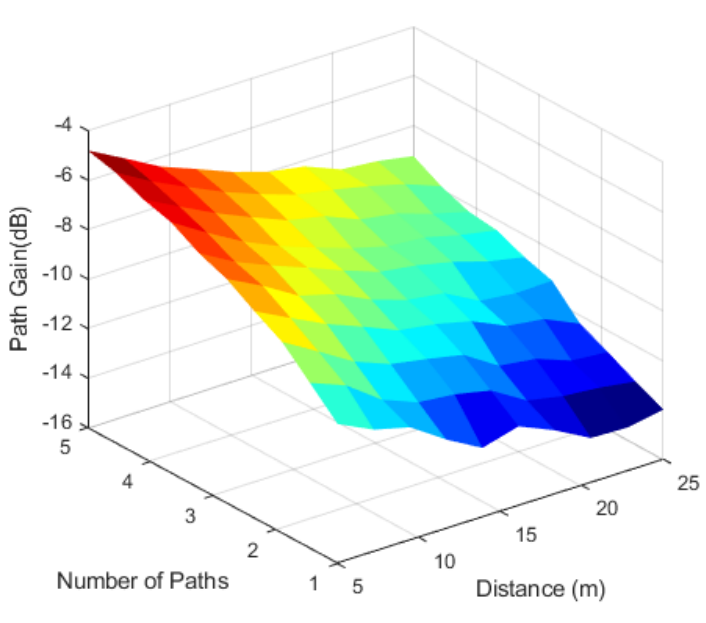}
    }
    \hfil
    \subfloat[Relationship between channel variables and environmental features in scenario 2.]{
        \includegraphics[width=0.23\linewidth]{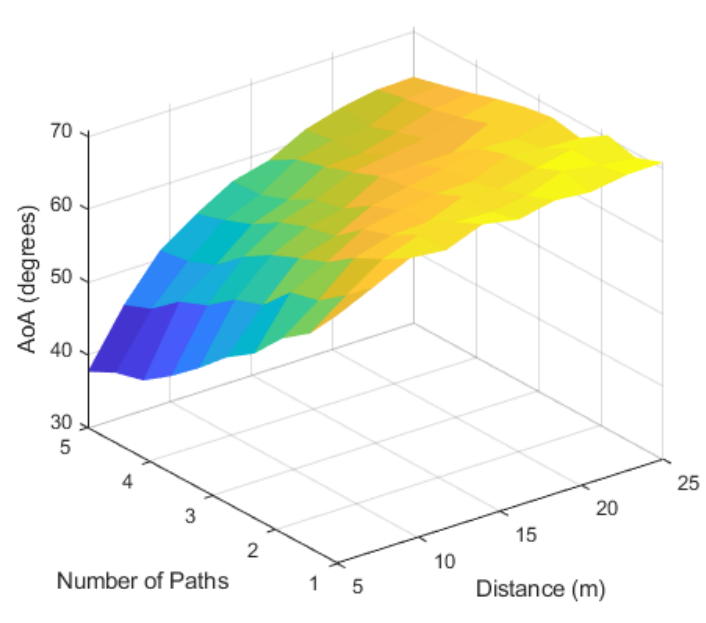}
        \includegraphics[width=0.23\linewidth]{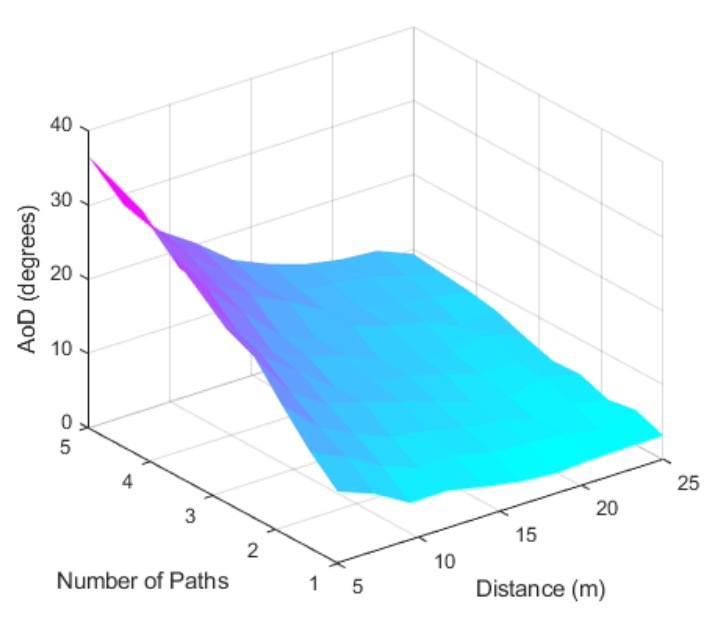}
        \includegraphics[width=0.23\linewidth]{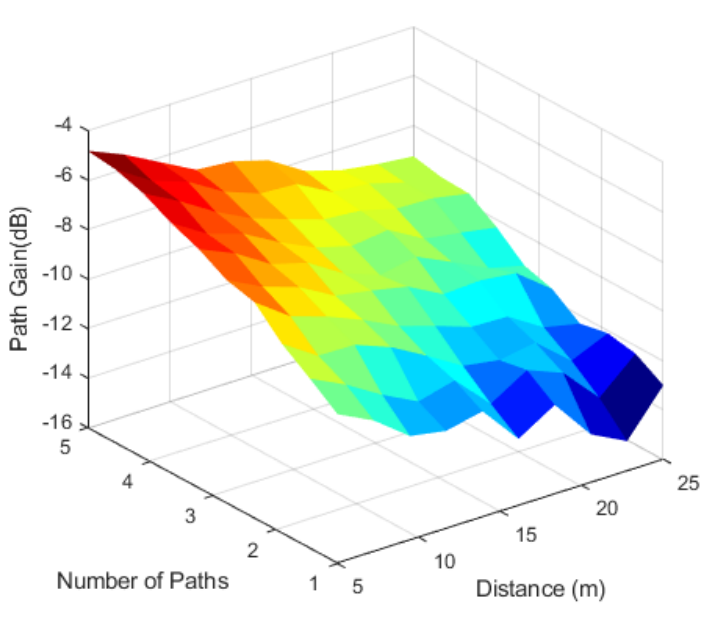}
    }
    \hfil
    \subfloat[Relationship between channel variables and environmental features in scenario 3.]{
        \includegraphics[width=0.23\linewidth]{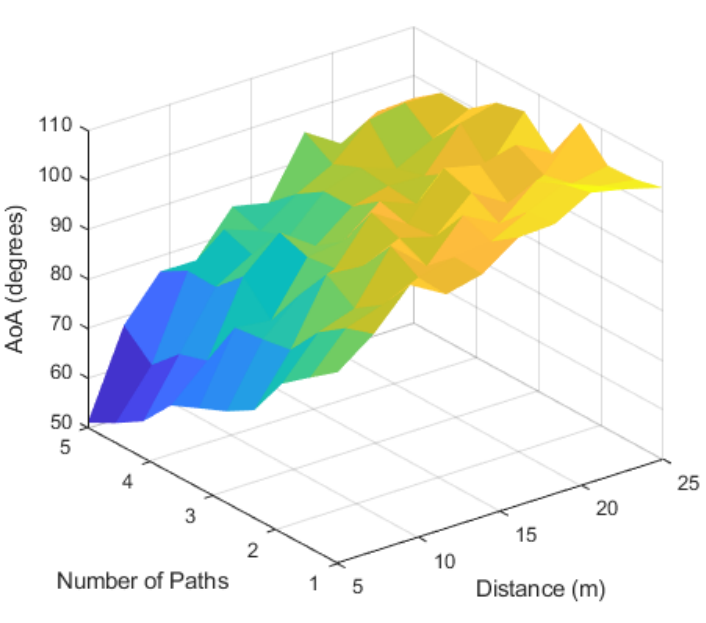}
        \includegraphics[width=0.23\linewidth]{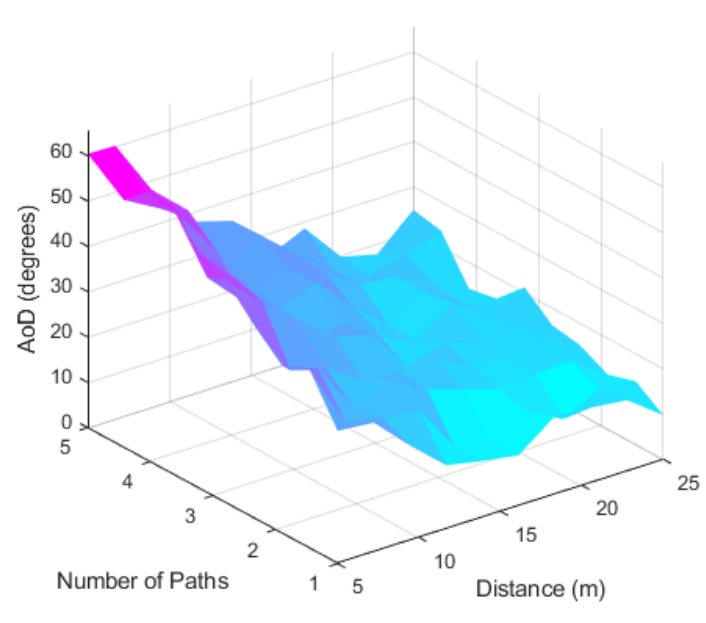}
        \includegraphics[width=0.23\linewidth]{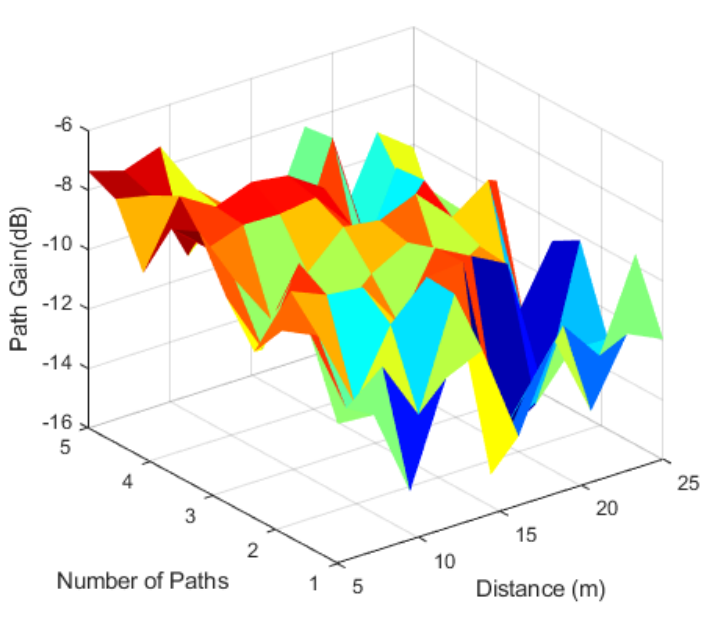}
    }
    \hfil
    \subfloat[Relationship between channel variables and environmental features in scenario 4.]{
        \includegraphics[width=0.23\linewidth]{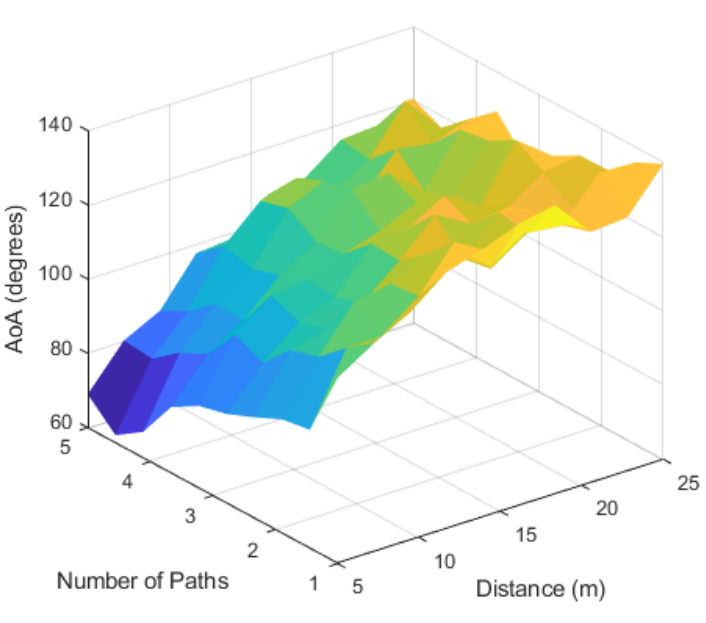}
        \includegraphics[width=0.23\linewidth]{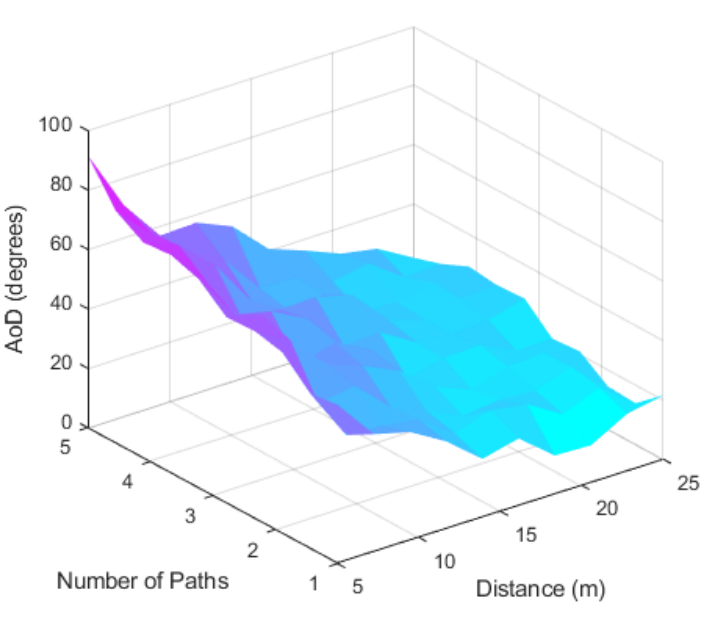}
        \includegraphics[width=0.23\linewidth]{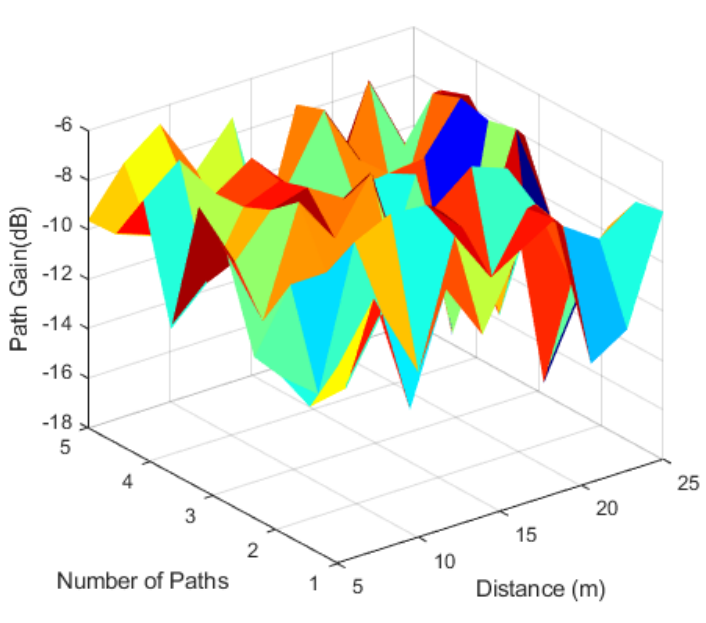}
    }
    \caption{Relationship between channel variables (AoA, AoD, and path gain) and environmental features across different scenarios.}
    \label{fig:channel_variables_scenarios}
\end{figure*}

\begin{figure}[t!]
    \centering
    \includegraphics[width=0.8\linewidth]{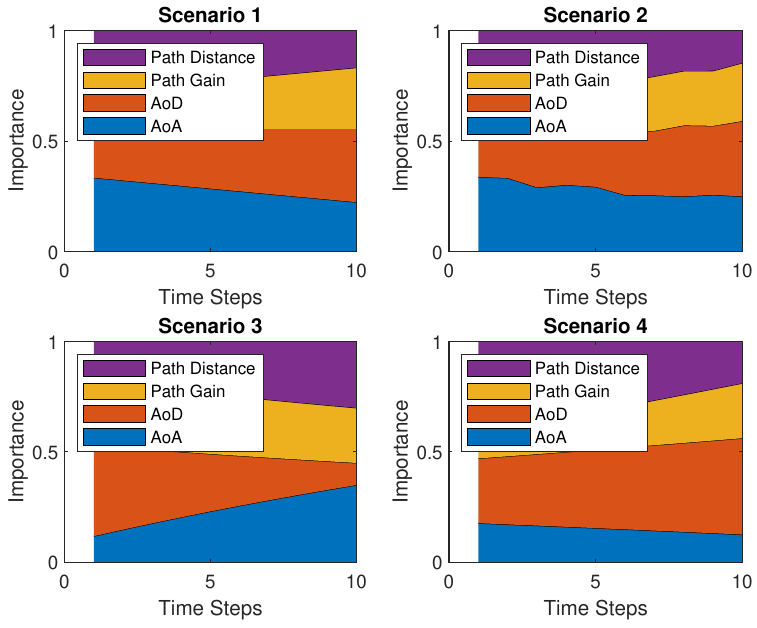}
    \caption{Importance of channel parameter over time in the scenarios.}
    \label{fig:importance_over_time}
\end{figure}

\begin{figure}[t!]
    \centering
    \includegraphics[width=0.9\linewidth]{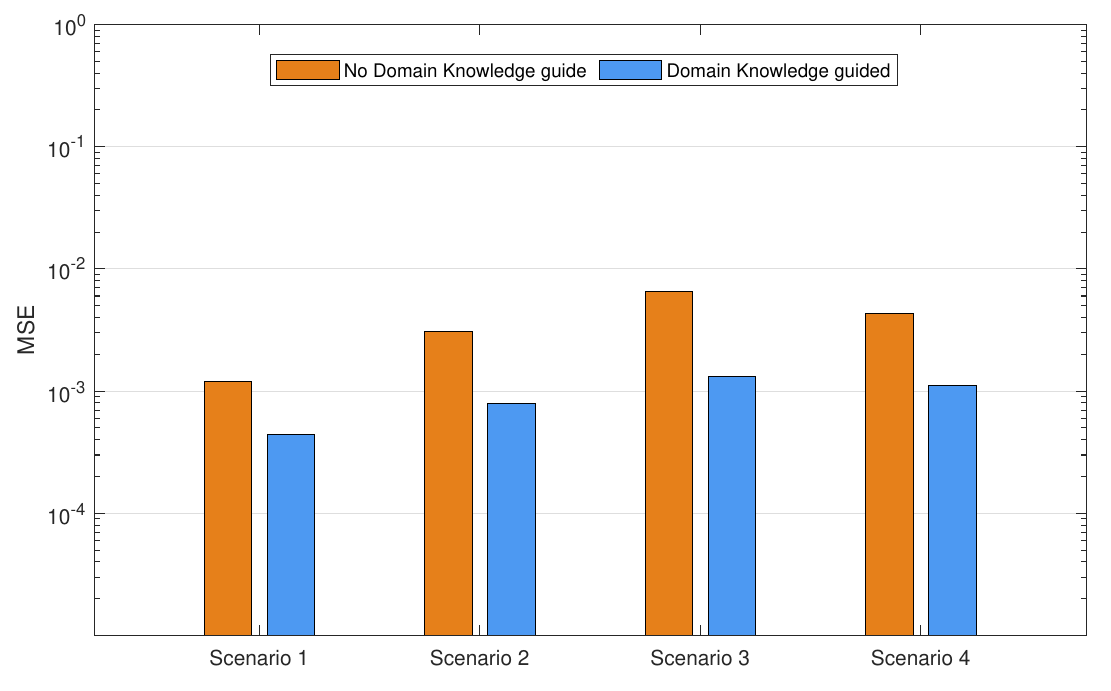}
    \caption{No domain guide vs.\ domain-guided approach.}
    \label{fig:domain_guided_vs_no}
\end{figure}

Fig. 6 presents a counterfactual analysis, where the model's performance is evaluated under different environmental conditions to examine how well it generalizes beyond the training scenario. All models, including the baselines and the proposed method, are trained on data from scenario 1.

Fig. 6 (a) compares the performance of the proposed and baseline methods in Scenario 1. The performance gap between VCD and other models is smaller than in the other results in Fig. 5. Nonetheless, the Multi-task + VCD model still achieves the lowest MSE outperforming other models by up to 95.59\% compared to the CNN-based method and 73.29\% compared to the Multi-task + GNN method. For MC, it remains the worst among all baselines even in scenario 1; because it is pilot-only and completes the grid with a low-rank fit, it smooths angle/delay structure and misses per-path changes, so its MSE stays highest. Notably, the GNN-based model shows relatively close performance to VCD. This may be because GNN captures spatial dependencies and structural relationships between objects, which can indirectly reflect some patterns that resemble causal interactions. However, VCD demonstrates significantly better performance than GNN because it effectively learns the causal relationships of hard-to-detect channel variable changes in complex environments like multi-path settings and thereby maintains high performance across various conditions. The mulT structure enables VCD to learn and more accurately reflect the complex interactions between channel characteristics and environmental factors.


Fig. 6(b) shows the performance of each model in the same environment as scenario 1 but with more vehicles. The increase in vehicles slightly reduces channel estimation accuracy, but the overall performance remains similar to that in Fig. 5(a). In this scenario, the Multi-task + VCD model achieves the lowest MSE and outperforms the other models. For MC, adding more vehicles makes it degrade the most among baselines: extra reflections and small Doppler shifts appear, but with a fixed pilot budget and low-rank filling, MC fails to capture these new paths and its error grows the fastest. As the number of vehicles increases, the multipath and reflection paths become more complex. Multi-task learning allows VCD to learn how these changes affect channel characteristics and incorporate them into its causal structure. It adapts to variations in channel variables caused by changes in position and speed and maintains prediction accuracy. The Multi-task + VCD model improves channel estimation accuracy by up to 88.39\% compared to the CNN-based method and 49.08\% compared to the Multi-task + GNN method. This improvement is achieved despite the increased environmental complexity, where the additional vehicles introduce more unpredictable multipath effects and reflections that make accurate channel estimation more challenging.



Fig. 6(c) and Fig. 6(d) comprehensively show the performance of each model in entirely different environments from the original training data. These scenarios are designed to evaluate how each model performs under various physical conditions different from those in the training data. Even in these new environments the Multi-task + VCD model achieves the lowest MSE and demonstrates high generalization ability compared to other models. For MC under domain shift (scenarios 3–4), it shows the largest error among baselines; unseen layouts change the angle/delay distribution and effective rank, so nuclear-norm completion over-smooths energy and mis-fills missing entries, widening its gap to GNN/VCD. The Multi-task + VCD model improves channel estimation accuracy by up to 99.9\% compared to the CNN-based method and 68.70\% compared to the Multi-task + GNN method. The performance gap between the proposed method and CNN becomes even larger in these scenarios because the proposed method effectively adapts to new environments by leveraging its causal learning structure to model the influence of environmental factors on channel characteristics. The GNN-based model also performs relatively well as it captures spatial dependencies and object interactions which helps it model some level of structural relationships in the environment. However, VCD extends beyond these learned relationships in new environments and captures complex channel characteristics such as multipath and reflection more effectively. These results show that VCD learns causal relationships from previous data and consistently predicts channel characteristics across diverse environmental conditions.


The impact of environmental factors on channel characteristics is examined across four distinct scenarios, as illustrated in Fig. 7. Each subfigure highlights the influence of the number of signal propagation paths on channel variables. 

Fig. 7(a) presents a sparse intersection(scenario 1) with minimal obstructions, resulting in a limited number of propagation paths. In this setting, the AoA increases gradually with distance, while the available signal paths remain relatively stable, leading to a smooth decline in path gain.

Fig. 7(b) illustrates a high density of moving vehicles(scenario 2) is introduced while maintaining the same structural environment as in Scenario 1. These vehicles create additional reflective surfaces, which increase the number of paths and cause fluctuations in both AoA and AoD, thereby intensifying multipath effects..

Fig. 7(c) presents a dense urban environment(scenario 3) characterized by frequent obstructions and multiple reflection points. Compared to Scenario 1, this setting exhibits a significantly larger number of paths, which leads to a wider AoA distribution and more unpredictable variations in path gain due to strong scattering and diffraction.

Fig. 7(d) represents an open road with moving traffic(scenario 4). In contrast to the dense urban scenario, the number of paths is lower, and fluctuations in AoA and AoD are less pronounced. However, reflections from open spaces introduce erratic variations in path gain.

Fig. 8 shows the relative importance of each channel variable during channel estimation over time across different scenarios, showing how the significance of each variable shifts with environmental changes. Path gain and distance are relatively more critical variables in scenarios 1 and 2. In contrast, in more dynamic environments such as scenarios 3 and 4, with greater complexity in signal paths, the importance of AoA and AoD increases. This adaptation reflects the model's ability to capture the causal relationships between environmental conditions and channel variables and demonstrates its effectiveness in modeling these variations. By understanding how environmental factors influence each channel variable, the model moves beyond simple statistical patterns, dynamically adjusting variable importance in response to environmental shifts. This adaptability supports the model's ability to maintain robust performance in highly dynamic conditions. 

These results (Fig. 7 and Fig. 8) illustrate how changes in the scenario alter the relationship between channel variables and environmental factors. This variation makes it difficult for conventional learning-based models to generalize across different environments. Traditional statistical models and machine learning-based approaches struggle in this setting because they rely on fixed patterns learned from training data. When faced with unseen scenarios, these models often fail to capture the shifting dependencies between channel and environmental variables. Even existing causal approaches \cite{annadani2021variational} are insufficient as they do not adapt quickly to these shifts.

The causal relationships identified by VCD reflect well-known physical principles, but explicitly modeling them is essential when applying AI to complex wireless networks. Direct data-driven learning often falls short in environments with dynamic and high-dimensional interactions. By integrating causal reasoning, VCD allows AI models to generalize across diverse scenarios and ensures accurate channel estimation under previously unseen conditions.

As shown in Fig. 9, the proposed ablation study compares two training strategies of our backbone: a no domain guided driven discovery and a domain-guided causal structure. In the training environment (scenario~1), both approaches achieve comparable accuracy, which implies that data-driven learning can fit the observed dependencies within the training set. However, in unseen environments (scenarios~2--4), the data-driven variant suffers from severe performance degradation, whereas the domain-guided model maintains significantly lower MSE. These results highlight that embedding physical priors into the causal graph not only provides interpretability but also substantially improves generalization to unseen propagation scenarios.

The proposed Multi-task + VCD framework achieves the lowest MSE in all scenarios. The reduction in estimation error lowers pilot overhead, which frees more resources for data transmission and increases throughput and spectrum utilization. Accurate estimation also improves beam control reliability, and in RIS-assisted MIMO it enables more effective RIS phase control, which increases coverage extension, spectral efficiency, and energy efficiency. The method is not tied to a specific dataset, and it uncovers causal relations between environmental and channel variables and ensures generalization across different scenarios and channel models. In addition, not only environmental factors but also various factors that influence channel variables can be considered by the framework. It can include antenna-centric effects such as beamwidth, array geometry, element spacing, radiation pattern, polarization, and mutual coupling as causal variables, and it can model their interactions with environmental factors within the same causal graph. This extension improves estimation fidelity, supports interpretation that aligns with physical constraints, and enables robust and resource-efficient operation in 5G and 6G THz networks.

\section{Conclusion}
In this paper, we have introduced a vision-based and causal neural network approach for channel estimation in urban THz communications. This method captures the relationship between physical features and wireless signal propagation. Through extensive simulations, we have demonstrated that the proposed model achieves up to twice the accuracy of conventional machine learning methods, such as CNN and LSTM-based channel estimators, and maintains strong performance in dynamic urban settings. We have shown that the model generalizes well to unseen environments and is a promising solution for 6G wireless networks. We have also incorporated causal reasoning to enhance robustness by accounting for complex indirect signal paths. This framework is expected to improve THz communication performance and enable more accurate and reliable channel estimates for 6G applications such as reconfigurable intelligent surfaces (RIS) and beamforming.

\bibliographystyle{IEEEtran}
\bibliography{References}

\vfill
\end{document}